%% file: main.tex
\documentclass[10pt,journal,compsoc]{IEEEtran}
\ifCLASSOPTIONcompsoc
  \usepackage[nocompress]{cite}
\else
  \usepackage{cite}
\fi
\usepackage{amsmath,amssymb,amsfonts}
\usepackage{algorithmic}
\usepackage{graphicx}
\usepackage{textcomp}
\usepackage{booktabs}
\usepackage{amssymb}
\usepackage{pifont}
\usepackage{multirow}
\usepackage{hyperref}
\usepackage[table,xcdraw]{xcolor}
\usepackage{enumitem}
\usepackage{threeparttable}
\usepackage{todonotes}
\usepackage{url}
\usepackage{makecell}
\usepackage{caption}
\usepackage{subcaption}
\usepackage{tikz}
\usepackage{circledsteps}
\usepackage{xcolor}
\usepackage{listings}
\usepackage{color}

\definecolor{dkgreen}{rgb}{0,0.6,0}
\definecolor{gray}{rgb}{0.5,0.5,0.5}
\definecolor{mauve}{rgb}{0.58,0,0.82}

\newcommand{\cmark}{\ding{51}}%
\newcommand{\xmark}{\ding{55}}%
\begin{document}

\title{Investigating Black-Box Function Recognition Using Hardware Performance Counters}

\author{Carlton~Shepherd, Benjamin~Semal, and Konstantinos~Markantonakis
\IEEEcompsocitemizethanks{\IEEEcompsocthanksitem All authors are of Royal Holloway, University of London, Egham, Surrey, United Kingdom. 
E-mail: carlton@linux.com.}}
\markboth{}%
{Shepherd \MakeLowercase{\textit{et al.}}}

\IEEEtitleabstractindextext{%
\begin{abstract}
This paper presents new methods and results for recognising black-box program functions using hardware performance counters (HPC), where an investigator can invoke and measure function calls. Important use cases include analysing compiled libraries, e.g.\ static and dynamic link libraries, and trusted execution environment (TEE) applications. We develop a generic approach to classify a comprehensive set of hardware events, e.g.\ branch mis-predictions and instruction retirements, to recognise standard benchmarking and cryptographic library functions. This includes various signing, verification and hash functions, and ciphers in numerous modes of operation. Three architectures are evaluated using off-the-shelf Intel/X86-64, ARM, and RISC-V CPUs. Next, we show that several known CVE-numbered OpenSSL vulnerabilities can be detected using HPC differences between patched and unpatched library versions. Further, we demonstrate that standardised cryptographic functions within ARM TrustZone TEE applications can be recognised using non-secure world HPC measurements, applying to platforms that insecurely perturb the performance monitoring unit (PMU) during TEE execution. High accuracy was achieved in all cases (86.22--99.83\%) depending on the application, architectural, and compilation assumptions. Lastly, we discuss mitigations, outstanding challenges, and directions for future research.
\end{abstract}

}

\maketitle
\input{sections/intro}

\input{sections/background}

\input{sections/cross-architectural-analysis}

\input{sections/openssl-vulnerabilities}
\input{sections/tee}
\input{sections/evaluation}
\input{sections/conc}

\ifCLASSOPTIONcompsoc
  \section*{Acknowledgments}
\else
  \section*{Acknowledgment}
\fi

This work received funding from the European Union's Horizon 2020 research and innovation programme under grant agreement No.\ 883156 (EXFILES). 

\bibliographystyle{IEEEtran}
\bibliography{bibliography}

\vspace{-1cm}
\begin{IEEEbiography}[{\includegraphics[width=1in,height=1.25in,clip,keepaspectratio]{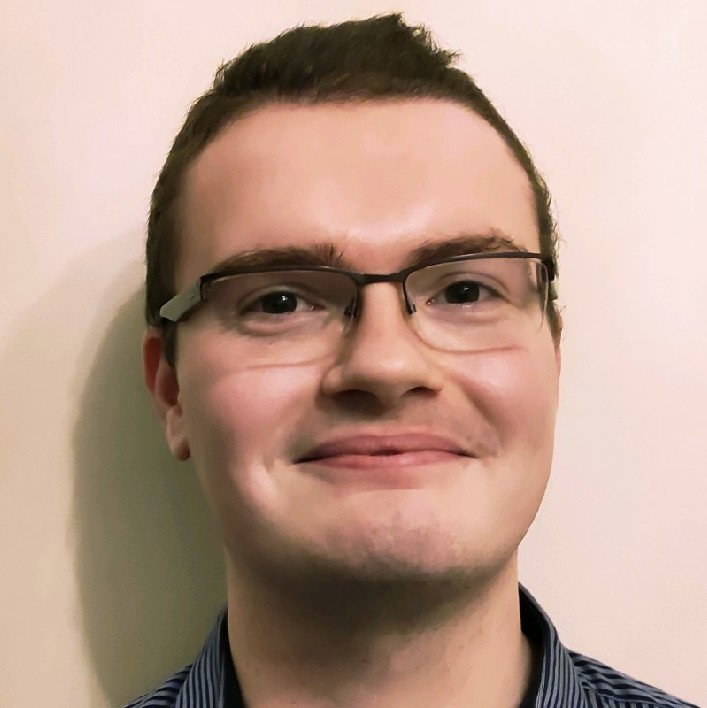}}]{Carlton Shepherd}  received his
Ph.D. in Information Security from the Information Security Group at Royal Holloway, University of London, U.K., and his B.S. in Computer Science from Newcastle University, U.K. He is currently a Research Fellow at the Information Security Group at Royal Holloway, University of London, where his research interests centre around the security of trusted execution environments (TEEs), security-enhanced CPU designs, and embedded systems security.
\end{IEEEbiography}
\vspace{-1cm}
\begin{IEEEbiography}[{\includegraphics[width=1in,height=1.25in,clip,keepaspectratio]{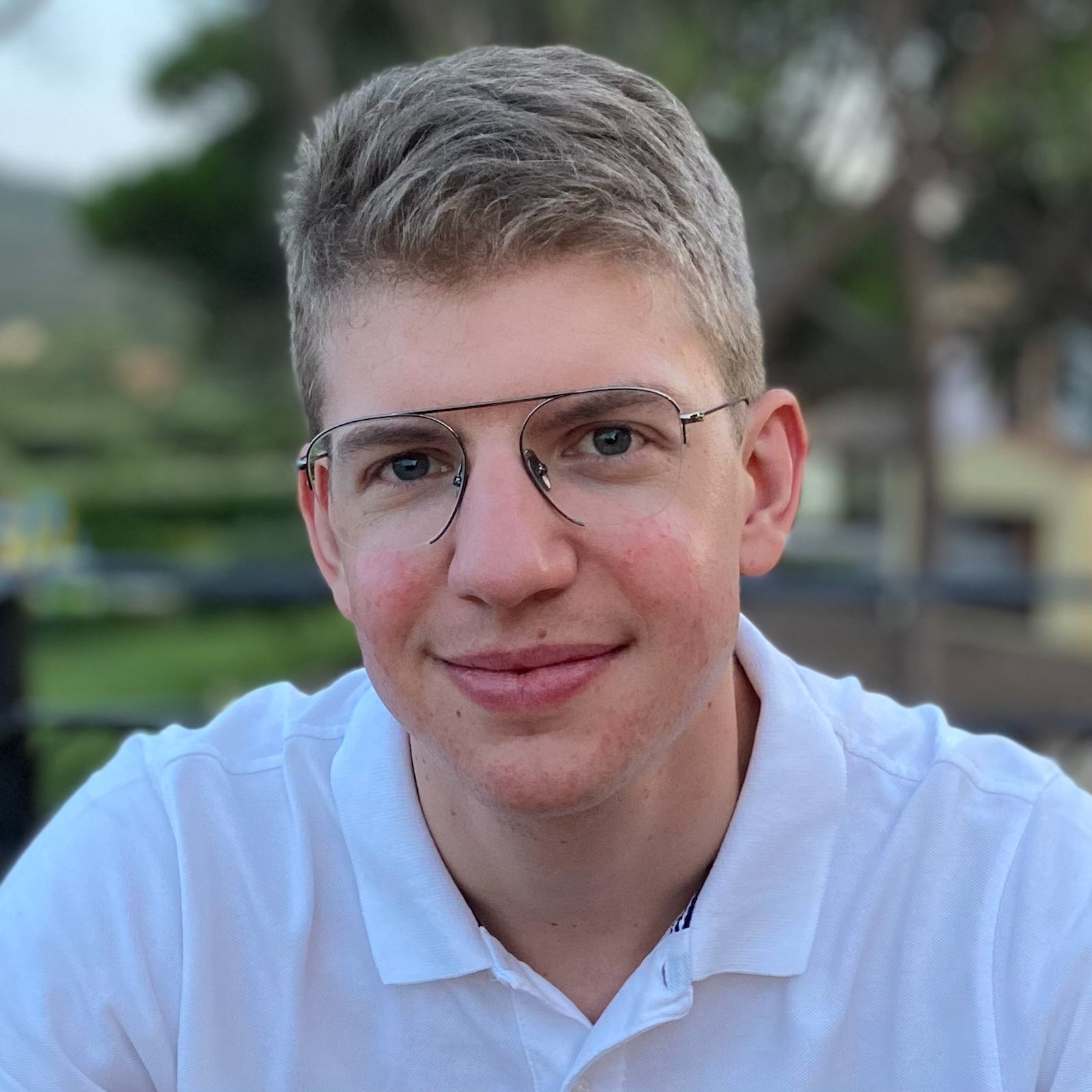}}]{Benjamin Semal}
received his M.Eng. in Electrical Engineering from Ecole Polytechnique Universitaire of Montpellier, France and M.S. in robotics from Cranfield University, U.K. He then worked as a hardware security analyst at UL Transaction Security. He later joined Royal Holloway, University of London to pursue a Ph.D.\ in Information Security. His research focusses on side-channel attacks for information leakage in multi-tenant environments. Benjamin now works as a security engineer at SERMA Security \& Safety evaluating point-of-sale devices and cryptographic modules.
\end{IEEEbiography}
\vspace{-1cm}
\begin{IEEEbiography}[{\includegraphics[width=1in,height=1.25in,clip,keepaspectratio]{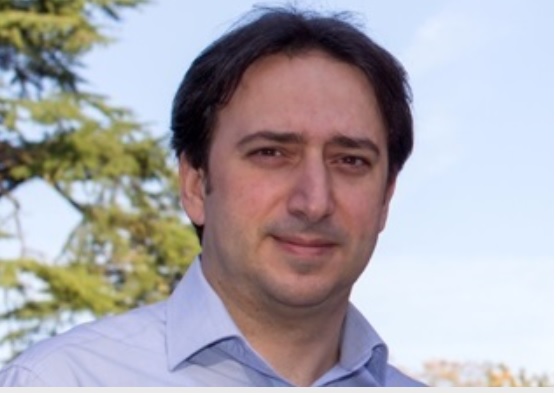}}]{Konstantinos Markantonakis}
received his B.S. in Computer Science from Lancaster University, U.K.; and his M.S.\ and Ph.D.\ in Information Security, and M.B.A.\ in International Management from Royal Holloway, University of London, London, U.K. He is currently the Director of the Smart Card and IoT Security Centre. He has co-authored over 190 papers in international conferences and journals. His research interests include smart card security, trusted execution environments, and the Internet of Things.
\end{IEEEbiography}

\clearpage

\appendices

\section{ARM and RISC-V Confusion Matrices}
\label{sec:other_confusion_matrices}

\begin{figure}
        \centering
        \includegraphics[width=1.04\linewidth,trim={0.7cm 0.7cm 0 0},clip]{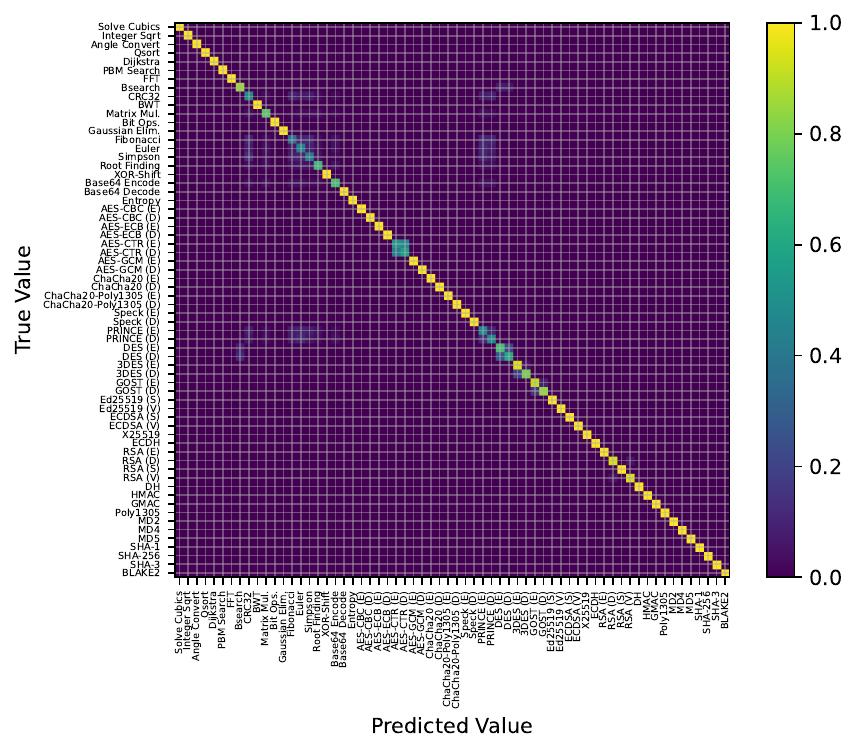}
        \caption{Normalised ARM confusion matrix.}
        \label{fig:classify_arm}
\end{figure}

\begin{figure}[t]
    \centering
    \begin{subfigure}{\linewidth}
         \centering
         \includegraphics[width=\textwidth,trim={0.7cm 0.7cm 1.8cm 0},clip]{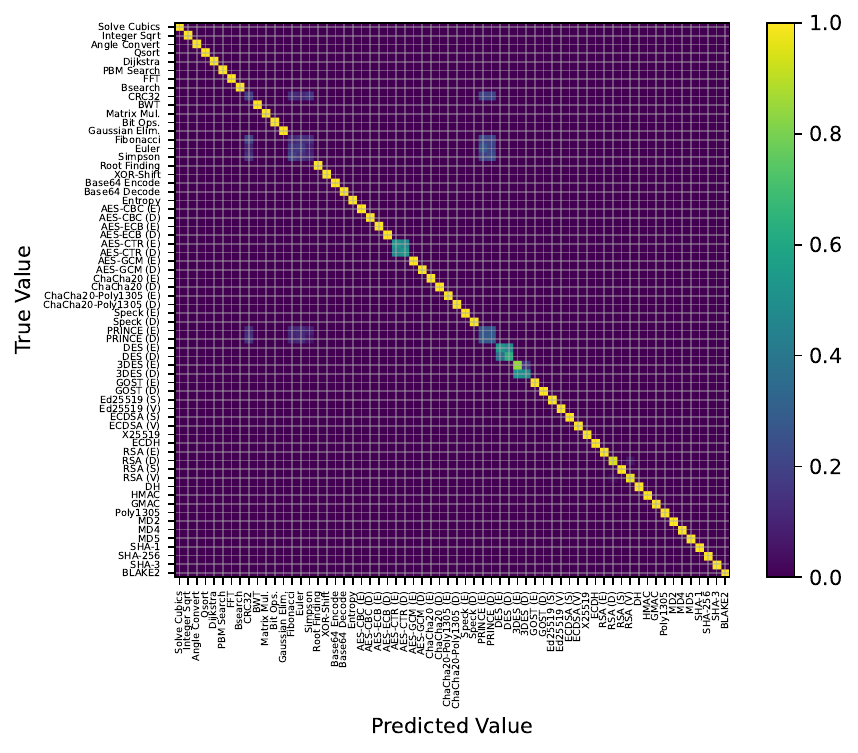}
         \caption{RISC-V (Privileged).}
         \label{fig:classify_riscv}
    \end{subfigure}
    \begin{subfigure}{\linewidth}
         \centering
         \includegraphics[width=\textwidth,trim={0.7cm 0.7cm 1.8cm 0},clip]{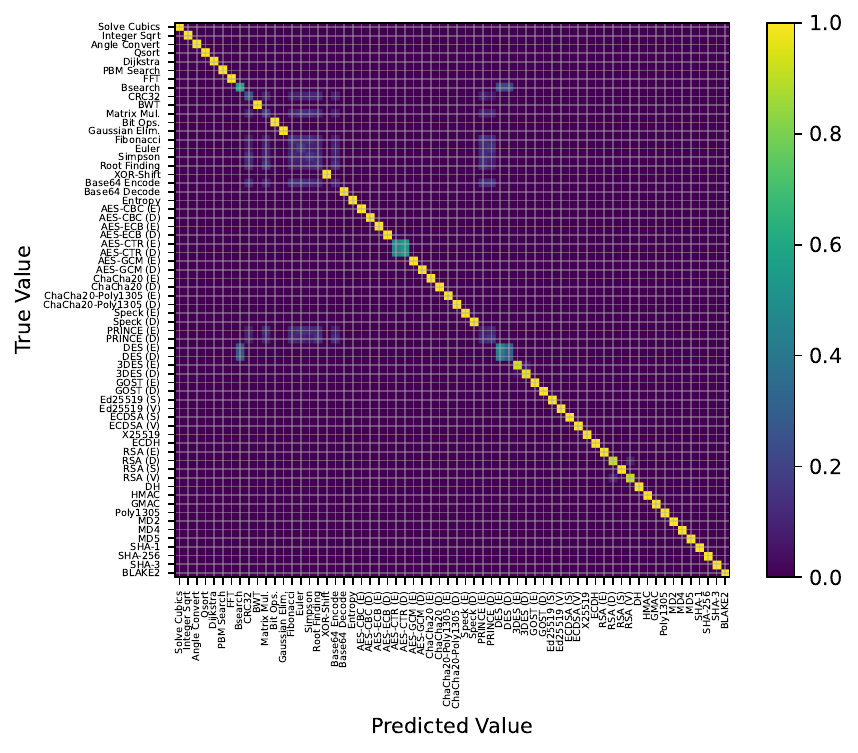}
         \caption{RISC-V (Unprivileged).}
         \label{fig:classify_riscv}
    \end{subfigure}
    \caption{Normalised RISC-V confusion matrices.}
    \label{fig:more_confusion_matrices}
\end{figure}

Confusion matrices relevant to \S\ref{sec:classification_results} for ARM and RISC-V function recognition are given in Figs.~\ref{fig:classify_arm} and \ref{fig:more_confusion_matrices} respectively.


\section{OpenSSL CVE Descriptions}
\label{sec:cve_description}

This appendix provides an overview of the six OpenSSL CVEs explored in \S\ref{sec:openssl_vulns} as follows:

\begin{itemize}
    \item[\textbf{(V1):}] \texttt{CVE-2018-5407}: \emph{``ECC scalar multiplication, used in e.g. ECDSA and ECDH, has been shown to be vulnerable to a microarchitecture timing side channel attack. An attacker with sufficient access to mount local timing attacks during ECDSA signature generation could recover the private key.''} Fixed in OpenSSL v1.1.0i. 
    \item[\textbf{(V2):}] \texttt{CVE-2018-0734}: \emph{``DSA signature algorithm has been shown to be vulnerable to a timing side channel attack. An attacker could use variations in the signing algorithm to recover the private key.''}. Fixed in v1.1.1a.
    \item[\textbf{(V3):}] \texttt{CVE-2018-0735}: \emph{``The OpenSSL ECDSA signature algorithm has been shown to be vulnerable to a timing side channel attack. An attacker could use variations in the signing algorithm to recover the private key.''} Fixed in v1.1.0j.
    \item[\textbf{(V4):}] \texttt{CVE-2018-0737}: \emph{``RSA key generation algorithm has been shown to be vulnerable to a cache timing side channel attack. An attacker with sufficient access to mount cache timing attacks during the RSA key generation process could recover the private key.''} Fixed in v1.1.0i.
    \item[\textbf{(V5):}] \texttt{CVE-2016-2178}: \emph{``Operations in the DSA signing algorithm should run in constant time in order to avoid side channel attacks. A flaw in the OpenSSL DSA implementation means that a non-constant time codepath is followed for certain operations. This has been demonstrated through a cache-timing attack to be sufficient for an attacker to recover the private DSA key.''} Fixed in v1.0.2i.
\item[\textbf{(V6):}] \texttt{CVE-2016-0702}: \emph{``A side-channel attack was found which makes use of cache-bank conflicts on the Intel Sandy-Bridge microarchitecture which could lead to the recovery of RSA keys. The ability to exploit this issue is limited as it relies on an attacker who has control of code in a thread running on the same hyper-threaded core as the victim thread which is performing decryptions.''} Fixed in v1.0.2g.
\end{itemize}

\newpage

\section{Model Hyper-parameter Details}
\label{app:hyperparams}

The model hyper-parameters used in the results in Tables \ref{tab:x86_classification_results}--\ref{tab:openssl-results-arm} and \ref{tab:ta_classification_accuracy} were derived from the ranges presented in Table~\ref{tab:model_hyperparams}.
\begin{table}[h]
\caption{Evaluated model hyper-parameters (Scikit-Learn nomenclature~\cite{scikit-learn}; default values used otherwise).}
\label{tab:model_hyperparams}
\begin{tabular}{@{}rl@{}}
\toprule
\textbf{Method} & \textbf{Hyper-parmeter ranges} \\ \midrule
\textbf{NB}     & Variance smoothing: $[\text{1e-1, 1e-3, 1e-5, 1e-7, 1e-9, 1e-11, 1e-13}]$\vspace{0.15cm}\\
\textbf{LR}     & $C$ (regularisation): $[\text{1e-4, 1e-3, 1e-2, 1e-1, 1e0, 1e1, 1e2, 1e3, 1e4}]$ \\
                & Optimisation solver: $[\text{'newton-cg', 'lbfgs', 'liblinear', 'sag', 'saga'}]$ \vspace{0.15cm}\\
\textbf{kNN}    & $N$ neighbours: $[\text{1, 3, 5, 10, 25, 50, 75, 100}]$ \vspace{0.15cm}\\
\textbf{DT}     & Split criterion: $[\text{'gini', 'log\_loss', 'entropy'}]$  \\
                & Max.\ tree depth: $[\text{2, 5, 7, 10, 25, 50, 75, 100, None}]$\vspace{0.15cm}\\
\textbf{RF}     &  Split criterion: $[\text{'gini', 'log\_loss', 'entropy'}]$  \\
                & Max.\ tree depth: $[\text{2, 5, 7, 10, 25, 50, 75, 100, None}]$ \\
                & $N$ estimators: $[\text{5, 10, 25, 50, 100, 250, 375, 500}]$\vspace{0.15cm}\\
\textbf{GBM}    & Learning rate: $[\text{0.01, 0.05, 0.1, 0.25, 0.5, 0.75, 1.0}]$\\
                & $N$ estimators: $[\text{5, 10, 25, 50, 100, 250, 375, 500}]$ \\
                & Max.\ tree depth: $[\text{1, 3, 5, 7, 10, 25, 50, 75, 100}]$ \vspace{0.15cm}\\
\textbf{LDA}    &   Optimisation solver: $[\text{'svd', 'lsqr', 'eigen'}]$ \vspace{0.15cm}\\
\textbf{SVM}    & Kernel: $[\text{'linear', 'poly', 'rbf', 'sigmoid'}]$ \\
                & $C$ (regularisation): $[\text{1e-4, 1e-3, 1e-2, 1e-1, 1e0, 1e1, 1e2, 1e3, 1e4}]$ \\ 
                & $\gamma$ (kernel coefficient): $[\text{'auto', 'scale'}]$ \vspace{0.15cm}\\
\textbf{MLP}    & Hidden layer dimensions: $[\text{(50,), (100,), (250,), (50,50), (100,100),}]$ \\
                & $\text{(250,250), (50,50,50), (100,100,100), (250,250,250), (100,250,100)}$ \\
                & $\text{(50,100,250,250), (50,50,50,50), (100,100,100,100),(250,250,250,250)}]$ \\
                & Activation functions: $[\text{'identity', 'logistic', 'tanh', 'relu'}]$\\\bottomrule
\end{tabular}
\end{table}

\end{document}

%% file: sections/intro.tex
\section{Introduction}
\IEEEPARstart{M}{odern} central processing units (CPUs) support a range of hardware performance counters (HPCs) for monitoring run-time memory accesses, pipeline events (e.g.\ instruction retired), cache hits, clock cycles, and more. Today's CPUs may support very few HPCs---under 10 on constrained microcontroller units---to over 100 events on Intel and AMD server chips~\cite{intel:sys_guide,amd:sys_guide}.  Originally intended for optimisation and debugging purposes, HPC events have found a myriad of security applications. For example, measurement sources for cache attacks~\cite{uhsadel2008exploiting}; intrusion detection~\cite{payer2016hexpads,yuan2011security}; malware detection~\cite{zhou2018hardware,singh2017detection,alam2019ratafia}; maintaining control-flow integrity~\cite{xia2012cfimon}; and reverse engineering proprietary CPU features~\cite{maurice2015reverse,helm2020reliable}. While the security implications of high-resolution HPCs have been acknowledged~\cite{riscv:unprivileged,arm:cortexa53_manual,trustedfirmware}, they remain widely available on commercial platforms.

In this paper, we explore a novel application where measurements are analysed \emph{en masse} from multiple counters in order to identify executed program functions. A generic supervised learning workflow is developed, where target functions are classified using events collected before and after their invocation.  Analysing exposed functions in this way can help ameliorate time-consuming binary patching and reverse-engineering, e.g.\ to implement precise code triggers, which is a major challenge in related work~\cite{shepherd:physical_fias_scas}.

To this end, we present the results of a three-part investigation. Firstly, \S\ref{sec:algorithm_identification} presents a foundational study on the feasibility of identifying functions from only HPC measurements, using a standard benchmarking suite (MiBench~\cite{guthaus2001mibench}) and four popular cryptography libraries (WolfSSL, Intel's Tinycrypt, Monocypher, and LibTomCrypt). Our approach identifies functions with 48.29\%--83.81\% accuracy (unprivileged execution) and 86.22\%--97.33\% (privileged), depending on the target architecture (X86-64/Intel, ARM Cortex-A, and RISC-V). Moreover, we analyse correlations of HPCs and use model inspection techniques to gauge their relative importance. From this, we distill a reduced set of the most effective counters for facilitating generalisation to other platforms that do not support a wide range of HPCs.

After this, \S\ref{sec:openssl_vulns} explores an offensive use case for detecting patches of known vulnerabilities. This has applications as a reconnaissance method during security evaluations, where we show how several OpenSSL (\texttt{libcrypto}) micro-architectural vulnerabilities can be recognised with 0.889--0.998 $F_1$-score (89.58\%--99.83\% accuracy). Next, \S\ref{sec:tee_reverse} investigates how a malicious spy process may use HPC measurements to recognise cryptographic algorithms executed by a trusted application (TA) within an ARM TrustZone-based trusted execution environment (TEE). Using OP-TEE---an open-source GlobalPlatform-compliant TEE---and a comprehensive set of GlobalPlatform TEE Client API~\cite{gp:tee} functions, the spy can recognise victim TA functions with 95.50\% accuracy. Finally, \S\ref{sec:evaluation} presents a security analysis, mitigations, and problems for future research.

\subsection{Threat Model} 
\label{sec:threat_model}

We consider an attacker, $A$, that aims to identify particular algorithms under execution given only high-level function calls and limited knowledge of its implementation.  This is often the case when analysing software with no debug symbols, function/variable names, and optimisation and code obfuscation techniques that inhibit readability. An example is the analysis of pre-existing compiled shared libraries on a target system. For instance, Windows dynamic link libraries (DLLs) and Linux shared objects, where the source code is inaccessible but where a spy application may link to and call functions of interest. A second example is TEE applications, e.g.\ ARM TrustZone TAs, where only high-level functions calls are exposed to untrusted world software~\cite{gp:tee}. $A$'s aim is to recognise functions from only CPU HPC measurements taken prior to and following their invocation. This requires kernel-mode code execution to access a full range of HPCs; however, we also explore cases where only user-mode counters are used. $A$ is also assumed to possess oracle access for collecting HPC measurements from idempotent functions, without restrictions on the number of permitted invocations.  In our approach, we use a model trained on known HPC-function mappings, i.e.\ on another system under $A$'s control, which is used for identifying unpatched functions, insecure cryptographic functions, and other tasks on the target.

\subsection{Contributions}

This paper presents the following contributions:
\begin{itemize}
    \item The design and evaluation of a generic approach for function recognition using HPC measurements. We develop a test-bed of standard cryptographic and non-cryptographic algorithms taken from widely used cryptographic libraries and MiBench~\cite{guthaus2001mibench} using three off-the-shelf RISC-V, ARM and X86-64 (Intel) platforms. Different privilege levels and compiler optimisations on overall performance are also examined. A feature importance analysis is conducted for determining a strong minimal set of HPCs towards facilitating generalisation.
    \item Methods and results of two use cases: \Circled{1} detecting unpatched versions of OpenSSL for several CVE-numbered vulnerabilities; and \Circled{2} recognising GlobalPlatform cryptographic functions executing in OP-TEE, a GlobalPlatform-compliant ARM TrustZone TEE implementation. This is a \emph{passive} vector that is difficult to mitigate in software against privileged adversaries. For both, experimental results indicate that HPCs can effectively recognise target program functions. Our approach aims to offer an alternative to orthogonal methods for reverse-engineering and vulnerability detection, like those requiring physical access~\cite{robyns2020practical,wilt2020toward} and static analysis and symbolic execution~\cite{zhang2016detecting,chakraborty2021deep}.
\end{itemize}

%% file: sections/background.tex
\section{Background}
\label{sec:background}

This section discusses related literature and background information on CPU performance events.

\subsection{Related Work}

HPCs have been used in various security applications, particularly as precise measurement sources for micro-architectural side-channels, like speculative execution and cache timing attacks~\cite{uhsadel2008exploiting,alam2017performance,li2018online}.  A significant body of work has also studied HPCs for malware detection in static and online environments, including rootkit, cryptocurrency miner, and ransomware detection~\cite{zhou2018hardware,singh2017detection,alam2019ratafia}. A general approach instruments target binaries to acquire measurements from the CPU's performance monitoring unit (PMU), which are then used to build custom statistical and machine learning models from known malware/safeware samples.

For intrusion detection, Eunomia~\cite{yuan2011security} traps and analyses sensitive syscalls during suspicious program execution. A trusted monitor analyses the preceding HPC measurements and permits/denies the call to prevent code injection, return-to-libc, and return-oriented programming (ROP) attacks.  Xia et al.~\cite{xia2012cfimon} tackle address control-flow integrity using Intel's Branch Trace Store (BTS), an Intel PMU buffer for storing control transfer events (e.g.\ jumps, calls, returns, and exceptions). The work builds legal sets of target addresses offline, after which branch traces of suspect applications are compared at run-time.   Payer~\cite{payer2016hexpads} proposed \textsc{HexPADS}, which examines performance events of target processes for detecting Rowhammer, cache attacks, and cross-VM address space layout randomisation (ASLR) breakages.

Copos and Murthy~\cite{copos2015inputfinder} developed a fuzzer that uses HPCs to build valid inputs for closed binaries, where binaries are instrumented for measuring the instruction retirement HPC before executing the program under different inputs. Control flow changes are detected for valid inputs using differences in instruction counters. Spisak~\cite{spisak2016hardware} developed a kernel-mode rootkit family that uses PMU interrupts to trap system events, such as syscalls. Interestingly, it is shown that TrustZone TAs can perturb the PMU on some consumer devices. Malone et al.~\cite{malone2011hardware} investigated static and dynamic software integrity verification using install-time \emph{vs.}\ run-time HPC measurements. Measurements from six HPCs are presented using four test programs; however, performance results are not given using standard evaluation metrics.

For reverse engineering, Maurice et al.~\cite{maurice2015reverse} use per-slice PMU access counter measurements to determine the cache slice assigned to a last-level cache (LLC) complex address for enabling cross-core LLC cache attacks. Helm et al.~\cite{helm2020reliable} use HPCs for understanding Intel's proprietary DRAM mapping mechanism for translating physical addresses to physical memory channels, banks, and ranks. Similarly, the work uses differences in per-channel PMU transfer counters while accessing different known physical addresses. 

 Physical attacks have been explored in work with similar applications but under different attack models. Robyns et al.~\cite{robyns2020practical} use a convolutional neural network (CNN) to detect eight cryptographic operations using electromagnetic (EM) emissions from a NodeMCU microcontroller, achieving 96\% accuracy. Wilt et al.~\cite{wilt2020toward} use a similar CNN-based approach for OS and malware detection with $\sim$84--99\% accuracy using EM radiation. Moreover, static analysis and symbolic execution have attracted significant attention for vulnerability detection, achieving 84\%--100\% accuracy under laboratory conditions, albeit with well-studied problems of model generalisability and state space explosion~\cite{chakraborty2021deep,zhang2016detecting}.

\subsection{Performance Monitoring Units (PMUs)}

Performance counters are available on all major CPU architectures within hardware PMUs, enabling the collection of detailed events with negligible overhead. HPCs are configured and accessed through special-purpose registers that update during execution. While different CPU architectures may count the same types of events, their availability and accessibility can differ significantly. We briefly describe the mechanics of PMUs on X86-64, ARM, and RISC-V. 

\subsubsection{X86}

The Intel PMU, introduced on Pentium CPUs, provides non-configurable registers for tracking fixed events and several programmable registers per logical core~\cite{intel:sys_guide}.  Intel Core CPUs support four general-purpose programmable registers and three fixed-function registers tracking elapsed core cycles, reference cycles, and retired instructions. Programmable registers are configured to simultaneously monitor one of $>$100 performance events on the Intel Xeon and Core architectures, such as branch mis-predictions and hits at various cache hierarchy levels. PMUs implement configuration and counter registers as model-specific registers (MSRs), accessible using the \texttt{RDMSR} and \texttt{WRMSR} instructions in ring 0 (kernel mode). The \texttt{RDPMC} instruction may also be used for accessing PMU counters at lower privilege levels if the \texttt{CR4.PCE} control register bit is set.  By default, ring 3 (user mode) access is granted to monotonic time-stamp counters (TSC) in a 64-bit register using the \texttt{RDTSC} instruction.  In addition to precise event counting, event-based sampling is supported for triggering a performance monitoring interrupt after exceeding threshold value (e.g.\ after $n$ events). AMD CPUs have minimal functional differences to Intel implementations for counting particular events, but do support more programmable events (six \emph{vs.}\ four)~\cite{payer2016hexpads,amd:sys_guide,intel:sys_guide}.

\subsubsection{ARM}
\label{sec:arm_counters}

The ARM PMU is a ubiquitous extension for ARM Cortex-A, -M, and -R processors. It supports per-core monitoring of similar but generally fewer high-level events to X86 CPUs. The ARM Cortex-A53, for example, contains a smaller cache hierarchy with a shared L2 cache at the highest level, precluding the ability to report L3 instruction or data cache events~\cite{arm:cortexa53_manual}. Like X86, fixed-function cycle counters are commonplace, and 2--8 general-purpose counters can be programmed to monitor the events using the \texttt{MRS} and \texttt{MSR} instructions (AArch64).   PMU registers can be configured to be accessed at any privilege mode (exception level) using the performance monitors control register (PMCR). Typically only kernel mode (EL3) processes may access PMU registers by default. ARM PMUs may also assert \texttt{nPMUIRQ} interrupt signals, e.g.\ counter overflows, which can be routed to an interrupt controller for prioritisation and masking.

\subsubsection{RISC-V}
\label{sec:risc-v}

The RISC-V Privileged~\cite{riscv:privileged} and Unprivileged~\cite{riscv:unprivileged} ISA specifications define separate instructions for accessing HPCs in different privilege modes. The Unprivileged ISA specifies 32 64-bit instructions for per-core counters in user- and supervisor-modes (U- and S-mode) and fixed-function counters for cycle count (\texttt{RDCYCLE}), real-time clock (\texttt{RDTIME}), and instruction retirements (\texttt{RDINSTRET}). Control and status register (CSR) space is reserved for 29 vendor-specific 64-bit HPC registers (\texttt{HPCCOUNTER3}--\texttt{HPCCOUNTER31}). The Privileged ISA specifies analogous CSR registers (\texttt{MCYCLE}, \texttt{MTIME}, \texttt{MINSTRET}) accessible only in machine-mode (M-mode), with 32 64-bit registers (\texttt{MHPMCOUNTER3}--\texttt{MHPMCOUNTER31}) allocated for vendor-specific HPCs. Note that RISC-V embedded systems are expected to possess only M- or M- and U-modes~\cite{riscv:privileged}, while workstations and servers are expected to support S-mode and the coming hypervisor (H)-mode extensions~\cite{riscv:unprivileged}.

%% file: sections/cross-architectural-analysis.tex
\section{Function Recognition: A Preliminary Study on X86, ARM, and RISC-V}
\label{sec:algorithm_identification}

This section presents a foundational study on recognising a large range of standard functions using HPC measurements. We discuss the methodological approach and evaluated functions, prior to implementation challenges and results.

\subsection{Overview}

We assume two processes shown in Fig.~\ref{fig:high-level}: a \emph{Spy} controlled by the adversary, and a \emph{Victim} that exposes high-level functions. In practical cases, the \emph{Victim} will assume the form of a compiled static or shared library with which the \emph{Spy} is statically or dynamically linked. Our approach then follows two steps: \Circled{1} HPC measurement (feature) vectors corresponding to each function are assigned labels according to its identifier, which are used for building the model hypothesis. Next, \Circled{2} uses newly measured values and the model from \Circled{1} to infer the executed function. We evaluate this using three platform architectures:

\begin{figure}
    \centering
    \includegraphics[width=0.95\linewidth]{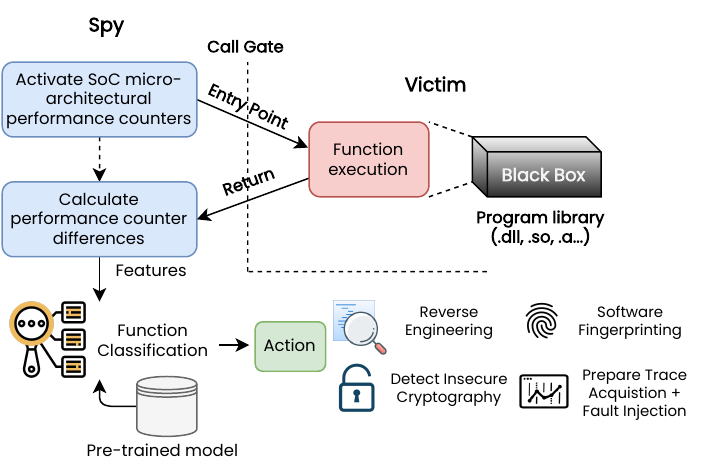}
    \caption{High-level approach.}
    \label{fig:high-level}
\end{figure}

\begin{itemize}
    \item \textbf{X86-64}. Dell Latitude 7410 with 8GB RAM and an Intel i5-10310U: 1.70GHz 64-bit quad-core, eight hyperthreads, and 256kB L1, 1MB L2, and 6MB L3 caches. Ubuntu 20.04 LTS was used with Linux kernel v5.12.
    \item \textbf{ARM}. Raspberry Pi 3B+ with 1GB SDRAM and a Broadcom BCM2837 system-on-chip (SoC): 1.4GHz 64-bit quad-core ARM Cortex-A53 CPU, with 32kB L1, 32kB L2, and no L3 cache. Raspbian OS was used, based on Debian 11/Bullseye, with Linux kernel v5.15.
    \item \textbf{RISC-V}. SiFive HiFive Rev.\ B with a FE310-G002 SoC: 320MHz 32-bit single-core CPU with RV32IMAC ISA support, 6kB L1 instruction cache, and 16kB SRAM. Supports privileged (M-) and unprivileged execution (U-mode). SiFive's Freedom E SDK was used as a hardware abstraction layer for application development.
\end{itemize}

We used PAPI~\cite{terpstra2010collecting} on our X86-64 and ARM devices, which provides portable HPC measurement acquisition using the Linux \emph{perf} subsystem. CPUs often expose extremely precise access to micro-architectural events, including pipeline- and DRAM controller-specific events, and proprietary features (e.g.\ Intel TSX), thereby preventing cross-platform compatibility.  To overcome this, PAPI implements micro-architectural dependent code and exposes common high-level HPC events, which we used for portability. Table \ref{tab:device-hpcs} shows the availability of these events on Intel, ARM and RISC-V. PAPI was compiled and dynamically linked with our benchmarking application that implemented the functions under test (\S\ref{sec:target_procedures}) on X86-64 and ARM. For collecting unprivileged events, our application executed in user-mode only. To access privileged HPCs, we set \texttt{kernel.perf\_event\_paranoid=-1} prior to execution to allow the reporting of kernel-mode counters in user space from Linux \emph{perf}. For RISC-V, the aforementioned assembly instructions were directly used for accessing HPC registers (\S\ref{sec:risc-v}) in M- (privileged) and U-modes (unprivileged).

\begin{table}
\centering
\caption{HPC events used on a per-device basis.}
\resizebox{\linewidth}{!}{%
\begin{threeparttable}
\begin{tabular}{@{}clccc@{}}
\toprule
 & & \multicolumn{3}{c}{\textbf{Device}} \\ 
\multirow{-2}{*}{\textbf{Method}} & \multicolumn{1}{c}{\multirow{-2}{*}{\textbf{Description}}} & \textbf{X86-64} & \textbf{ARM} & \textbf{RISC-V} \\ \midrule
\cellcolor[HTML]{bad4ff}RDTSCP & Cycle count since a reset. & \cellcolor[HTML]{C4FFAD}\cmark & \cellcolor[HTML]{FFA1A1}\xmark & \cellcolor[HTML]{FFA1A1}\xmark  \\
\cellcolor[HTML]{FFFFC7}L1\_DCM & L1 data cache misses. & \cellcolor[HTML]{C4FFAD}\cmark & \cellcolor[HTML]{C4FFAD}\cmark & \cellcolor[HTML]{FFA1A1}\xmark \\
\cellcolor[HTML]{FFFFC7}L1\_ICM & L1 instruction cache misses. & \cellcolor[HTML]{C4FFAD}\cmark  & \cellcolor[HTML]{C4FFAD}\cmark & \cellcolor[HTML]{C4FFAD}\cmark \\
\cellcolor[HTML]{FFFFC7}L1\_TCM & L1 total cache misses. & \cellcolor[HTML]{C4FFAD}\cmark & \cellcolor[HTML]{FFA1A1}\xmark &\cellcolor[HTML]{FFA1A1}\xmark  \\
\cellcolor[HTML]{FFFFC7}L1\_LDM & L1 load misses. & \cellcolor[HTML]{C4FFAD}\cmark & \cellcolor[HTML]{FFA1A1}\xmark &\cellcolor[HTML]{FFA1A1}\xmark  \\
\cellcolor[HTML]{FFFFC7}L1\_DCA & L1 data cache accesses. & \cellcolor[HTML]{FFA1A1}\xmark & \cellcolor[HTML]{C4FFAD}\cmark & \cellcolor[HTML]{FFA1A1}\xmark \\
\cellcolor[HTML]{FFFFC7}L1\_STM & L1 store misses. & \cellcolor[HTML]{C4FFAD}\cmark & \cellcolor[HTML]{FFA1A1}\xmark &\cellcolor[HTML]{FFA1A1}\xmark  \\
\cellcolor[HTML]{FFFFC7}L2\_DCM & L2 data cache misses. & \cellcolor[HTML]{C4FFAD}\cmark & \cellcolor[HTML]{C4FFAD}\cmark & \cellcolor[HTML]{FFA1A1}\xmark \\
\cellcolor[HTML]{FFFFC7}L2\_ICM & L2 instruction cache misses. & \cellcolor[HTML]{C4FFAD}\cmark & \cellcolor[HTML]{FFA1A1}\xmark & \cellcolor[HTML]{FFA1A1}\xmark \\
\cellcolor[HTML]{FFFFC7}L2\_TCM & L2 total cache misses. & \cellcolor[HTML]{C4FFAD}\cmark & \cellcolor[HTML]{FFA1A1}\xmark & \cellcolor[HTML]{FFA1A1}\xmark \\
\cellcolor[HTML]{FFFFC7}L2\_LDM & L2 load misses. & \cellcolor[HTML]{C4FFAD}\cmark & \cellcolor[HTML]{FFA1A1}\xmark & \cellcolor[HTML]{FFA1A1}\xmark \\
\cellcolor[HTML]{FFFFC7}L2\_DCR & L2 data cache reads. & \cellcolor[HTML]{C4FFAD}\cmark & \cellcolor[HTML]{FFA1A1}\xmark & \cellcolor[HTML]{FFA1A1}\xmark \\
\cellcolor[HTML]{FFFFC7}L2\_STM & L2 store misses. & \cellcolor[HTML]{C4FFAD}\cmark & \cellcolor[HTML]{FFA1A1}\xmark & \cellcolor[HTML]{FFA1A1}\xmark \\
\cellcolor[HTML]{FFFFC7}L2\_DCA & L2 data cache accesses. & \cellcolor[HTML]{C4FFAD}\cmark & \cellcolor[HTML]{C4FFAD}\cmark & \cellcolor[HTML]{FFA1A1}\xmark \\
\cellcolor[HTML]{FFFFC7}L2\_ICR & L2 instruction cache reads. & \cellcolor[HTML]{C4FFAD}\cmark & \cellcolor[HTML]{FFA1A1}\xmark & \cellcolor[HTML]{FFA1A1}\xmark \\
\cellcolor[HTML]{FFFFC7}L2\_ICH & L2 instruction cache hits. & \cellcolor[HTML]{C4FFAD}\cmark & \cellcolor[HTML]{FFA1A1}\xmark & \cellcolor[HTML]{FFA1A1}\xmark \\
\cellcolor[HTML]{FFFFC7}L2\_ICA & L2 instruction cache accesses. & \cellcolor[HTML]{C4FFAD}\cmark & \cellcolor[HTML]{FFA1A1}\xmark & \cellcolor[HTML]{FFA1A1}\xmark \\
\cellcolor[HTML]{FFFFC7}L2\_TCA & L2 total cache accesses. & \cellcolor[HTML]{C4FFAD}\cmark & \cellcolor[HTML]{FFA1A1}\xmark & \cellcolor[HTML]{FFA1A1}\xmark \\
\cellcolor[HTML]{FFFFC7}L2\_TCR & L2 total cache reads. & \cellcolor[HTML]{C4FFAD}\cmark & \cellcolor[HTML]{FFA1A1}\xmark & \cellcolor[HTML]{FFA1A1}\xmark \\
\cellcolor[HTML]{FFFFC7}L2\_TCW & L2 total cache writes. & \cellcolor[HTML]{C4FFAD}\cmark & \cellcolor[HTML]{FFA1A1}\xmark & \cellcolor[HTML]{FFA1A1}\xmark \\
\cellcolor[HTML]{FFFFC7}L3\_TCM & L3 total cache misses. & \cellcolor[HTML]{C4FFAD}\cmark & \cellcolor[HTML]{FFA1A1}\xmark &\cellcolor[HTML]{FFA1A1}\xmark  \\
\cellcolor[HTML]{FFFFC7}L3\_LDM & L3 load misses. & \cellcolor[HTML]{C4FFAD}\cmark & \cellcolor[HTML]{FFA1A1}\xmark &\cellcolor[HTML]{FFA1A1}\xmark  \\
\cellcolor[HTML]{FFFFC7}L3\_DCA & L3 data cache accesses. & \cellcolor[HTML]{C4FFAD}\cmark & \cellcolor[HTML]{FFA1A1}\xmark &\cellcolor[HTML]{FFA1A1}\xmark  \\
\cellcolor[HTML]{FFFFC7}L3\_DCR & L3 data cache reads. & \cellcolor[HTML]{C4FFAD}\cmark & \cellcolor[HTML]{FFA1A1}\xmark &\cellcolor[HTML]{FFA1A1}\xmark  \\
\cellcolor[HTML]{FFFFC7}L3\_DCW & L3 data cache writes. & \cellcolor[HTML]{C4FFAD}\cmark & \cellcolor[HTML]{FFA1A1}\xmark &\cellcolor[HTML]{FFA1A1}\xmark  \\
\cellcolor[HTML]{FFFFC7}L3\_ICA & L3 instruction cache accesses. & \cellcolor[HTML]{C4FFAD}\cmark & \cellcolor[HTML]{FFA1A1}\xmark &\cellcolor[HTML]{FFA1A1}\xmark  \\
\cellcolor[HTML]{FFFFC7}L3\_ICR & L3 instruction cache reads. & \cellcolor[HTML]{C4FFAD}\cmark & \cellcolor[HTML]{FFA1A1}\xmark &\cellcolor[HTML]{FFA1A1}\xmark  \\
\cellcolor[HTML]{FFFFC7}L3\_TCA & L3 total cache accesses. & \cellcolor[HTML]{C4FFAD}\cmark & \cellcolor[HTML]{FFA1A1}\xmark &\cellcolor[HTML]{FFA1A1}\xmark  \\
\cellcolor[HTML]{FFFFC7}L3\_TCR & L3 total cache reads. & \cellcolor[HTML]{C4FFAD}\cmark & \cellcolor[HTML]{FFA1A1}\xmark &\cellcolor[HTML]{FFA1A1}\xmark  \\
\cellcolor[HTML]{FFFFC7}L3\_TCW & L3 total cache writes. & \cellcolor[HTML]{C4FFAD}\cmark & \cellcolor[HTML]{FFA1A1}\xmark &\cellcolor[HTML]{FFA1A1}\xmark  \\
\cellcolor[HTML]{FFFFC7}CA\_SNP & Requests for a cache snoop. & \cellcolor[HTML]{C4FFAD}\cmark & \cellcolor[HTML]{FFA1A1}\xmark & \cellcolor[HTML]{FFA1A1}\xmark \\
\cellcolor[HTML]{FFFFC7}CA\_SHR & Requests for exclusive access to a shared cache line. & \cellcolor[HTML]{C4FFAD}\cmark & \cellcolor[HTML]{FFA1A1}\xmark & \cellcolor[HTML]{FFA1A1}\xmark \\
\cellcolor[HTML]{FFFFC7}CA\_CLN & Requests for exclusive access to a clean cache line. & \cellcolor[HTML]{C4FFAD}\cmark & \cellcolor[HTML]{FFA1A1}\xmark & \cellcolor[HTML]{FFA1A1}\xmark \\
\cellcolor[HTML]{FFFFC7}CA\_ITV & Requests for cache line intervention. & \cellcolor[HTML]{C4FFAD}\cmark & \cellcolor[HTML]{FFA1A1}\xmark & \cellcolor[HTML]{FFA1A1}\xmark\\
\cellcolor[HTML]{FFFFC7}TLB\_DM & Data TLB misses. & \cellcolor[HTML]{C4FFAD}\cmark & \cellcolor[HTML]{C4FFAD}\cmark & \cellcolor[HTML]{FFA1A1}\xmark \\
\cellcolor[HTML]{FFFFC7}TLB\_IM & Instruction TLB misses. & \cellcolor[HTML]{C4FFAD}\cmark & \cellcolor[HTML]{C4FFAD}\cmark & \cellcolor[HTML]{FFA1A1}\xmark \\
\cellcolor[HTML]{FFFFC7}PRF\_DM & Data pre-fetch cache misses. & \cellcolor[HTML]{C4FFAD}\cmark & \cellcolor[HTML]{FFA1A1}\xmark & \cellcolor[HTML]{FFA1A1}\xmark \\
\cellcolor[HTML]{FFFFC7}MEM\_WCY & Cycles stalled for memory writes. & \cellcolor[HTML]{C4FFAD}\cmark & \cellcolor[HTML]{FFA1A1}\xmark & \cellcolor[HTML]{FFA1A1}\xmark \\
\cellcolor[HTML]{FFFFC7}STL\_ICY & Cycles with no instruction issue. & \cellcolor[HTML]{C4FFAD}\cmark & \cellcolor[HTML]{FFA1A1}\xmark & \cellcolor[HTML]{FFA1A1}\xmark \\
\cellcolor[HTML]{FFFFC7}FUL\_ICY & Cycles with maximum instruction issue. & \cellcolor[HTML]{C4FFAD}\cmark & \cellcolor[HTML]{FFA1A1}\xmark & \cellcolor[HTML]{FFA1A1}\xmark \\
\cellcolor[HTML]{FFFFC7}BR\_UCN & Unconditional branch instructions. & \cellcolor[HTML]{C4FFAD}\cmark & \cellcolor[HTML]{FFA1A1}\xmark & \cellcolor[HTML]{FFA1A1}\xmark \\
\cellcolor[HTML]{FFFFC7}BR\_CN & Conditional branch instructions. & \cellcolor[HTML]{C4FFAD}\cmark & \cellcolor[HTML]{FFA1A1}\xmark & \cellcolor[HTML]{FFA1A1}\xmark \\
\cellcolor[HTML]{FFFFC7}BR\_TKN & Conditional branches taken. & \cellcolor[HTML]{C4FFAD}\cmark & \cellcolor[HTML]{FFA1A1}\xmark & \cellcolor[HTML]{FFA1A1}\xmark \\
\cellcolor[HTML]{FFFFC7}BR\_NTK & Conditional branch instructions not taken. & \cellcolor[HTML]{C4FFAD}\cmark & \cellcolor[HTML]{FFA1A1}\xmark & \cellcolor[HTML]{FFA1A1}\xmark \\
\cellcolor[HTML]{FFFFC7}BR\_MSP & Conditional branch mispredictions. & \cellcolor[HTML]{C4FFAD}\cmark & \cellcolor[HTML]{C4FFAD}\cmark & \cellcolor[HTML]{FFA1A1}\xmark \\
\cellcolor[HTML]{FFFFC7}BR\_PRC & Conditional branches correctly predicted. & \cellcolor[HTML]{C4FFAD}\cmark & \cellcolor[HTML]{FFA1A1}\xmark & \cellcolor[HTML]{FFA1A1}\xmark \\
\cellcolor[HTML]{FFFFC7}TOT\_INS & Instructions completed. & \cellcolor[HTML]{C4FFAD}\cmark & \cellcolor[HTML]{C4FFAD}\cmark  & \cellcolor[HTML]{FFA1A1}\xmark \\
\cellcolor[HTML]{FFFFC7}LD\_INS & Load instructions. & \cellcolor[HTML]{C4FFAD}\cmark & \cellcolor[HTML]{C4FFAD}\cmark & \cellcolor[HTML]{FFA1A1}\xmark \\
\cellcolor[HTML]{FFFFC7}SR\_INS & Store instructions. & \cellcolor[HTML]{C4FFAD}\cmark & \cellcolor[HTML]{C4FFAD}\cmark & \cellcolor[HTML]{FFA1A1}\xmark \\
\cellcolor[HTML]{FFFFC7}BR\_INS & Branch instructions. & \cellcolor[HTML]{C4FFAD}\cmark & \cellcolor[HTML]{C4FFAD}\cmark & \cellcolor[HTML]{FFA1A1}\xmark \\
\cellcolor[HTML]{FFFFC7}RES\_STL & Cycles stalled on any resource. & \cellcolor[HTML]{C4FFAD}\cmark & \cellcolor[HTML]{FFA1A1}\xmark & \cellcolor[HTML]{FFA1A1}\xmark  \\
\cellcolor[HTML]{FFFFC7}TOT\_CYC & Total cycles executed. & \cellcolor[HTML]{C4FFAD}\cmark & \cellcolor[HTML]{C4FFAD}\cmark  & \cellcolor[HTML]{FFA1A1}\xmark \\
\cellcolor[HTML]{FFFFC7}LST\_INS & Load and store instructions executed & \cellcolor[HTML]{C4FFAD}\cmark &\cellcolor[HTML]{FFA1A1}\xmark & \cellcolor[HTML]{FFA1A1}\xmark \\ 
\cellcolor[HTML]{FFFFC7}SP\_OPS & Single-precision floating point operations. & \cellcolor[HTML]{C4FFAD}\cmark &\cellcolor[HTML]{FFA1A1}\xmark & \cellcolor[HTML]{FFA1A1}\xmark \\ 
\cellcolor[HTML]{FFFFC7}DP\_OPS & Double-precision floating point operations. & \cellcolor[HTML]{C4FFAD}\cmark &\cellcolor[HTML]{FFA1A1}\xmark & \cellcolor[HTML]{FFA1A1}\xmark \\ 
\cellcolor[HTML]{FFFFC7}VEC\_SP & Single precision SIMD instructions. & \cellcolor[HTML]{C4FFAD}\cmark &\cellcolor[HTML]{FFA1A1}\xmark & \cellcolor[HTML]{FFA1A1}\xmark \\ 
\cellcolor[HTML]{FFFFC7}VEC\_DP & Double precision SIMD instructions. & \cellcolor[HTML]{C4FFAD}\cmark &\cellcolor[HTML]{FFA1A1}\xmark & \cellcolor[HTML]{FFA1A1}\xmark \\ 
\cellcolor[HTML]{bad4ff}RDCYCLE & U-mode reference cycle count. & \cellcolor[HTML]{FFA1A1}\xmark & \cellcolor[HTML]{FFA1A1}\xmark & \cellcolor[HTML]{C4FFAD}\cmark \\
\cellcolor[HTML]{bad4ff}RDINSTRET & U-mode instructions retired. & \cellcolor[HTML]{FFA1A1}\xmark & \cellcolor[HTML]{FFA1A1}\xmark & \cellcolor[HTML]{C4FFAD}\cmark \\
\cellcolor[HTML]{bad4ff}RDTIME & U-mode real-time counter & \cellcolor[HTML]{FFA1A1}\xmark & \cellcolor[HTML]{FFA1A1}\xmark & \cellcolor[HTML]{C4FFAD}\cmark \\
\cellcolor[HTML]{FFFFC7}MCYCLE & M-mode cycle counter & \cellcolor[HTML]{FFA1A1}\xmark & \cellcolor[HTML]{FFA1A1}\xmark & \cellcolor[HTML]{C4FFAD}\cmark \\
\cellcolor[HTML]{FFFFC7}MINSTRET & M-mode instructions retired. & \cellcolor[HTML]{FFA1A1}\xmark & \cellcolor[HTML]{FFA1A1}\xmark  & \cellcolor[HTML]{C4FFAD}\cmark \\
\cellcolor[HTML]{FFFFC7}MTIME & M-mode real-time counter. & \cellcolor[HTML]{FFA1A1}\xmark & \cellcolor[HTML]{FFA1A1}\xmark & \cellcolor[HTML]{C4FFAD}\cmark \\ \bottomrule
\end{tabular}
\begin{tablenotes}
\item X86-64: Intel i5-10310U; ARM: ARM Cortex-A53; RISC-V: SiFive FE310.
\item Blue cells indicate user-mode counters; yellow denote privileged counters.
\end{tablenotes}
\end{threeparttable}
}
\label{tab:device-hpcs}
\vspace{-0.1cm}
\end{table}

\subsection{Target Procedures}
\label{sec:target_procedures}

We developed a test-bed comprising 64 functions listed in Table~\ref{tab:test-bed_functions}, incorporating the MiBench benchmarking suite with common embedded systems procedures~\cite{guthaus2001mibench}, alongside modern cryptographic algorithms.  For the latter, we used an extensive range of common functions available through WolfSSL in addition to reference implementations of GOST, Speck, and PRINCE.   Our benchmarking tool used random input buffers in the range [8B, 32B, 256B, 512b, 1024kB, 2048kB, 4096kB] for encryption, signing, MAC, and hashing algorithms. This was used to avoid potential HPC measurement biases if only a fixed-size input was used for functions that accepted an arbitrary byte buffer and its length. Random keys were generated for all symmetric algorithms and keyed MACs (e.g.\ AES in all modes, ChaCha20, Prince, DES and HMAC).  Similarly, random public-private key pairs and secret and public values were generated for asymmetric and key exchange algorithms respectively (e.g.\ ECDSA, X25519, ECDH, and RSA). HPC measurements were not inclusive of this process. Fresh initialisation vectors (IVs), counter values and nonces were also generated where applicable (e.g.\ ChaCha20 and AES-CTR). RSA operations were split approximately equally using a random padding scheme and key length. RSAES-OAEP or RSAES-PKCS\#1v1.5 were used for encryption/decryption, and RSASSA-PSS or RSASSA-PKCS\#1v1.5 for signing/verification under 1024-, 2048-, 3072-, and 4096-bit key lengths. Similarly, ECDSA and ECDH used a random curve from P-256, P-384, and P-521, while AES used 128-, 196, and 256-bit key lengths.  For ease of implementation, corresponding inverse operations (e.g.\ verification for signing) were called with the same parameters immediately following the original operation.

\begin{table}[t]
\caption{Test-bed target functions.}
\label{tab:test-bed_functions}
\resizebox{\linewidth}{!}{%
\begin{threeparttable}
\begin{tabular}{@{}lll@{}}
\toprule
\multicolumn{3}{c}{\textbf{Non-cryptographic Functions}} \\ \midrule
1. Solve Cubics & 8. Bsearch  & 15. Euler \\
2. Integer Sqrt  & 9. CRC32 & 16. Simpson \\
3. Angle Convert & 10. BWT & 17. Root Finding \\
4. Qsort & 11. Matrix Mul.\ & 18. XOR-Shift \\
5. Dijkstra\ & 12. Bit Ops.\ & 19. Base64 Encode \\
6. PBM Search & 13. Gaussian Elim.\ & 20. Base64 Decode \\
7. FFT & 14. Fibonacci & 21. Entropy \\\midrule
\multicolumn{3}{c}{\textbf{Cryptographic Functions}} \\ \midrule
22-23. AES-ECB (E+D) & 40-41. 3DES (E+D) & 56. GMAC \\
24-25. AES-CBC (E+D) & 42-43. GOST (E+D) & 57. Poly1305  \\
26-27. AES-CTR (E+D) & 44-45. Ed25519 (S+V) & 58. MD2  \\
28-29. AES-GCM (E+D) & 46-47. ECDSA (S+V) & 59. MD4   \\
30-31. ChaCha20 (E+D) & 48. X25519 & 60. MD5 \\
32-33.  ChaCha20+Poly1305 (E+D)  & 49. ECDH & 61. SHA-1 \\
34-35. Speck (E+D)  & 50-53. RSA (E+D, S+V) & 62. SHA-256 \\
36-37. PRINCE (E+D)  & 54. DH  & 63. SHA-3 \\
38-39. DES (E+D) & 55. HMAC  & 64. BLAKE2 \\\bottomrule
\end{tabular}
\begin{tablenotes}
\item E+D: Encrypt and decrypt. S+V: Sign and verify.
\end{tablenotes}
\end{threeparttable}
}
\end{table}

\subsection{Methodology}
\label{sec:algorithm_identification_methodology}

For each available HPC event in Table~\ref{tab:device-hpcs}, we collected measurements from 10,000 invocations for each of the 64 test-bed functions. For instance, Qsort on RISC-V was invoked 70,000 times in total to collect the necessary measurements (10,000 invocations $\times$ 7 available HPCs). These $N$-dimensional feature vectors ($N=49$, X86-64; $N=13$, ARM; $N=7$, RISC-V) were mapped to the class label of the procedure ($[0, 64) \in \mathbb{N}$), i.e.\ $\sim$31.3M invocations in total (X86-64), 8.3M (ARM), and 4.4M (RISC-V).  We then formed two data sets: $\textbf{A}$ and $\textbf{B}$, representing HPCs available in unprivileged and privileged modes respectively.

Each data set was split into training and test sets using a 80:20 ratio, before applying Z-score normalisation to produce features with zero mean and unit variance. Given the relatively large and balanced many-class classification problem, conventional classification accuracy was used as the evaluation metric. A preliminary experiment initially explored a basic template approach using four similarity metrics---Euclidean distance, Minkowski distance, cross-correlation, and covariance---to classify test vectors against the training set. However, the approach produced unpromising results: $<$40\% accuracy in the best cases using branch predictions and total instructions. We thus resorted to training nine supervised classifiers: na\"{i}ve Bayes (NB), logistic regression (LR), linear discriminant analysis (LDA), decision tree (DT), gradient boosting machines (GBM), random forest (RF), support vector machine (SVM), and a multi-layer perceptron (MLP). Hyper-parameter optimisation was conducted using an exhaustive grid search and $k$-fold ($k=10$) cross-validation (CV), with the best performing CV classifier scoring the test set.

\subsection{Implementation Challenges}
\label{sec:hpc_implementation_challenges}

Using HPCs requires careful thought to avoid measurement bias and other pitfalls. Some phenomena have been investigated in related literature~\cite{das2019sok,weaver2013non}, while others have attracted little attention, like cache warming and compiler optimisations. We discuss these challenges forthwith.

\subsubsection{HPC Accuracy}

Run-to-run variations in HPC measurements is a challenge on commercial CPUs for program analysis~\cite{weaver2013non,das2019sok}. Counter measurements are not perfectly replicated between sequential program executions, with a 0.5\%--2\% deviation observed in counter values using standard benchmarks~\cite{weaver2013non}. This arises principally from:
\begin{itemize}
    \item \emph{Event non-determinism}. External events, particularly hardware interrupts and page faults, can cause small deviations in vulnerable HPC events, e.g.\ load and store counters, which are unpredictable and difficult to reproduce. Another source of variation is the pipeline effects arising, for instance, from out-of-order execution. This can also perturb absolute counter values, although the effects can be negligible if the monitoring of very short instruction sequences is avoided~\cite{arm:cortexa53_manual}. 
    \item \emph{Overcount}. CPU implementation differences and errata can overcount events. For example, instructions retired events can be overcounted by exceptions and pseudo-instructions where micro-coded instructions differ to the instructions that are actually executed. Specifically, X87/SSE exceptions, OS lazy floating point handling, and \texttt{fldcw}, \texttt{fldenv}, \texttt{frstor}, \texttt{maskmovq}, \texttt{emms}, \texttt{cvtpd2pi}, \texttt{cvttpd2pi}, \texttt{sfence}, \texttt{mfence} instructions are known overcount sources~\cite{weaver2013non}. 
\end{itemize}

The effects of non-determinism cannot be entirely eliminated. Rather, we model interactions from many individual measurements (millions of features) as a mitigation against single, run-to-run variations. This assumes the ability to measure large numbers of target functions idempotently, i.e.\ their behaviour changes insignificantly between invocations, and without restricting the number of permitted calls (see \S\ref{sec:threat_model}). Importantly, we do not rely upon exact measurements, but only that the deviations between different function executions confer enough discriminitive power for classification.  Moreover, we use measurements from \emph{multiple} counters to maximise the discriminative benefits of disparate event types, which also minimises potential issues (e.g.\ overcounting bias) of using any single counter. X86 CPUs can also measure the number of hardware interrupts---a known source of non-determinism---which was used as a feature (Table~\ref{tab:device-hpcs}).

\subsubsection{Function Implementation Variations}

It is important to discern between multiple implementations of the \emph{same} algorithm. Statistical models may fail to generalise when presented with HPC events of alternative libraries that, while similar algorithmically, contain code differences that can affect HPC measurements (e.g.\ API calling conventions). Considering this, we ported additional implementations from Monocypher (for ChaCha20, Poly1305, Ed25519, X25519, BLAKE2), Intel's Tinycrypt (AES, ECDSA, ECDH, HMAC, SHA-256), and LibTomCrypt (AES, MD2, MD4, MD5, SHA-1, SHA-256, 3/DES, RSA, DH, GMAC, HMAC) on our target platforms.\footnote{The choice of these libraries was to construct a common test-bed; very few cryptographic libraries (e.g.\ OpenSSL and GnuTLS) currently offer 32-bit RISC-V MCU support.} For these functions, measurements were taken by cycling through each implementation for each invocation, and mapping them to the same label. For instance, MD5 measurements from LibTomCrypt \emph{and} WolfSSL implementations were assigned the same label.

\subsubsection{Cache Warming}

Preliminary experiments showed that cache-based events, such as L1 and L2 misses, and their correlated values---discussed further in \S\ref{sec:correlation_analysis}---were markedly higher during initial function executions. This subsided after several executions per function in order to converge to stability ($<$3\% requiring an average of 9.7 executions, X86-64; 7.0, ARM; 7.3, RISC-V). We hypothesise that cache warming was partially responsible, which has not been in related literatur on HPC-based security applications. Within the context of this work, the reliance on large numbers of measurements renders this effect negligible ($<$0.5\%). However, the reader is warned of potential bias in constrained environments where only a limited number of measurements can be collected.

\subsubsection{Measuring Multiple Performance Events}

Today's PMUs support a small number of programmable counter registers, typically 4--8 depending on the architecture~\cite{intel:sys_guide,amd:sys_guide}. Consequently, measuring several events requires: \Circled{1} Using a single counter register per execution; \Circled{2} Batching HPC measurements, with the batch size equalling the number of supported counter registers; or \Circled{3} Using software multiplexing provided by some tools, e.g.\ PAPI, where counters are time-shared over many performance events. We used \Circled{1} as a conservative method at the cost of collection time. With \Circled{2}, collection time could be reduced by a constant factor (of the max.\ supported counters), while \Circled{3} minimises collection time at the cost of accuracy~\cite{terpstra2010collecting}.

\subsubsection{Compiler Optimisations} 
\label{sec:gcc}

Modern compilers improve performance and code size at the expense of compilation time and debuggability; for example, using loop unrolling, if-conversions, tail-call optimisation, inline expansion, and register allocation. This is an important consideration given the difficulty of knowing \emph{a priori} the exact compilation parameters of target software, e.g.\ a dynamic library. Some compiler optimisations can have a material effect on HPC measurements for the same program: Choi et al.~\cite{choi2001impact}, for instance, showed that if-conversions can reduce branch mis-predictions by 29\% on Intel CPUs, which would reflect in branch-related HPCs.

To address this, we compiled the test-bed and collected measurements from test devices under different GCC optimisation options: disabling optimisation (O0, the default setting); minimising code size (Os); and maximum optimisation (O3) that sacrifices compile time and memory usage for performance. The reader is referred to the GCC documentation for a comprehensive breakdown of the optimisations utilised with each flag~\cite{gcc:docs}. To emulate environments where compiler optimisations are unknown, a mixed data set was formed by concatenating and randomly shuffling measurement vectors collected under each flag.

\subsection{Results}
\label{sec:classification_results}

\begin{figure*}
    \centering
    \begin{subfigure}{0.535\linewidth}
        \centering
        \hspace{-1.4em}
        \includegraphics[width=1.04\linewidth,trim={0.7cm 0.7cm 0 0},clip]{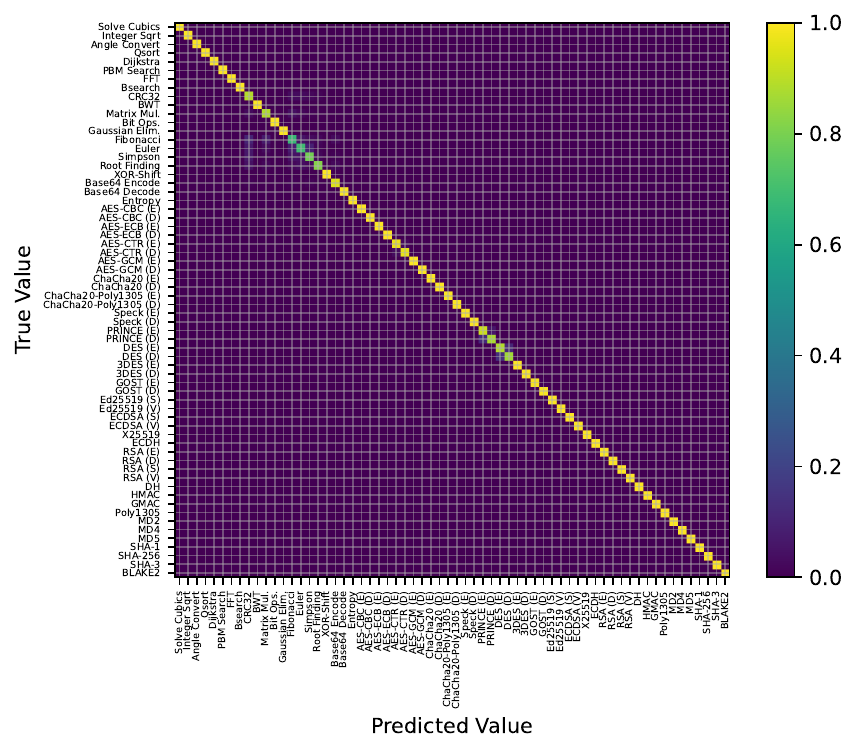}
        \caption{Privileged HPCs.}
        \label{fig:classify_x86_priv}
    \end{subfigure}
    \begin{subfigure}{0.46\linewidth}
         \centering
         \hspace{-1em}
         \includegraphics[width=1.035\linewidth,trim={0.7cm 0.7cm 2cm 0},clip]{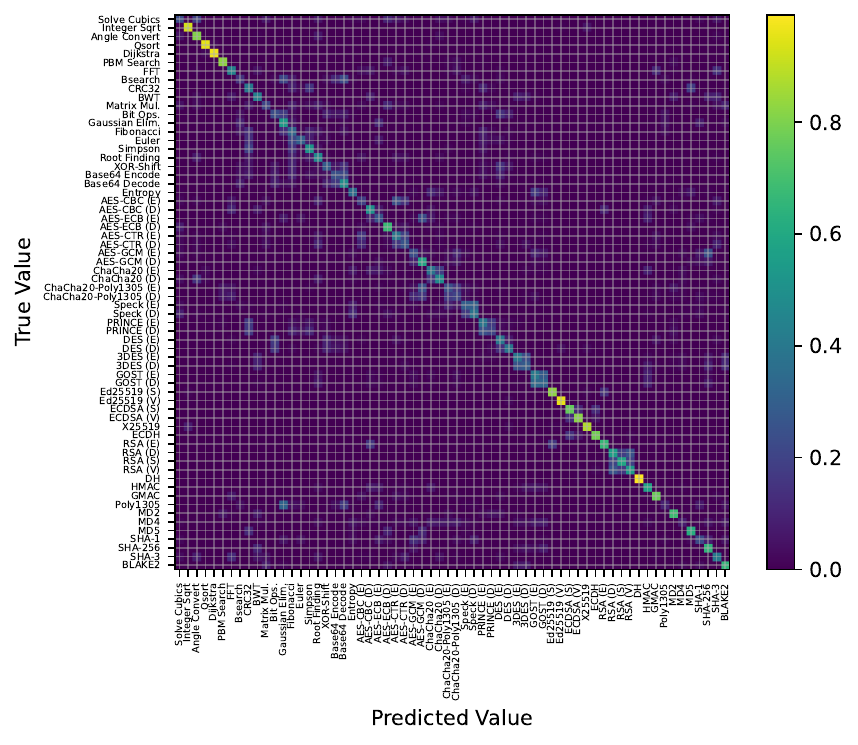}
         \caption{Unprivileged HPCs.}
         \label{fig:classify_x86_unpriv}
    \end{subfigure}
    \caption{Normalised confusion matrices for X86-64 (X-axis: Predicted value, Y-axis: True value).}
    \label{fig:confusion_matrix_x86}
\end{figure*}

\begin{table}[t]
\centering
\caption{Classification accuracy (Intel i5-10310U; in \%; data set A = unprivileged HPCs, B = privileged).}
\label{tab:x86_classification_results}
\resizebox{\linewidth}{!}{%
\begin{tabular}{@{}crccccccccc@{}}
\toprule
 & \multicolumn{1}{c}{\textbf{}} & \multicolumn{9}{c}{\textbf{Classifier}} \\ \cmidrule(l){3-11} 
\multicolumn{2}{l}{\textbf{Data set}} & \textbf{NB} & \textbf{LR} & \textbf{kNN} & \textbf{DT} & \textbf{RF} & \textbf{GBM} & \textbf{LDA} & \textbf{SVM} & \textbf{MLP} \\ \midrule
 & \textbf{O0} &  49.30 &   50.24 & 60.24   & 59.20  & \textbf{66.31}  & 65.79   &  61.56  &  54.44 &  51.83 \\
\multirow{2}{*}{\textbf{A}} & \textbf{Os} &   49.22 &  50.63  & 58.30  & 64.46 & 66.07  &  \textbf{66.56} & 61.87  & 62.43  &  54.62   \\
 & \textbf{O3} &  48.48   &  49.30 & 59.08 & \textbf{60.27}  & 59.81  & 49.71  &  35.57 & 45.46 &  47.28 \\
 & \textbf{Mixed} &  37.21 & 32.80  & \textbf{48.29}  & 48.03 &  47.67  &    40.52          &   26.30           &     45.39         &     37.02  \\ \midrule
& \textbf{O0} &  98.95 &  96.47  & 89.25  & 99.53 & \textbf{99.70}  & 99.02 & 94.31  & 99.04  &  86.55 \\
\multirow{2}{*}{\textbf{B}} & \textbf{Os} & 93.03 & 89.27 & 87.71 & 95.89 & 95.99 & \textbf{97.49} & 88.29 & 92.35 & 90.06 \\
 & \textbf{O3} &   87.55 &  86.08  &  82.75 & 92.01 &  \textbf{98.51} &  83.78 &  74.44 & 80.53  & 84.60 \\
 & \textbf{Mixed} & 84.78 & 73.95 & 85.14 & 95.40 & \textbf{97.33} & 91.01 & 80.86 & 83.78 & 85.95 \\ \bottomrule
\end{tabular}

}
\end{table}
\begin{table}[t]
\centering
\caption{Classification accuracy (ARM Cortex-A53; privileged counters).}
\label{tab:arm_classification_results}
\resizebox{\linewidth}{!}{%
\begin{tabular}{@{}crccccccccc@{}}
\toprule
 & \multicolumn{1}{c}{\textbf{}} & \multicolumn{9}{c}{\textbf{Classifier}} \\ \cmidrule(l){3-11} 
\multicolumn{2}{r}{\textbf{Data set}} & \textbf{NB} & \textbf{LR} & \textbf{kNN} & \textbf{DT} & \textbf{RF} & \textbf{GBM} & \textbf{LDA} & \textbf{SVM} & \textbf{MLP} \\ \midrule
 & \textbf{O0} &  85.65 &   87.27 & 82.34   & 96.87  & 96.74  & \textbf{96.81}   &  80.42  &  83.65 &  84.01  \\
 & \textbf{Os} &   77.91 &  77.45  & 82.70  & 90.31 & 88.59  &  \textbf{91.32} & 86.28  & 84.13  &  84.30   \\
 & \textbf{O3} &  79.21   &  82.99 & 82.85 & 88.90  & \textbf{91.40}  & 89.43  &  74.37 & 87.62 &  74.82 \\
 & \textbf{Mixed} &  73.97 & 70.08  & 80.35  & 88.96 & 88.99  &    \textbf{90.68}         &   74.00          &     80.37         &     69.09  \\ \bottomrule
\end{tabular}
}
\end{table}
\begin{table}[t!]
\centering
\caption{Classification accuracy (SiFive E31 SoC; data set A = unprivileged HPCs, B = privileged).}
\label{tab:riscv_classification_results}
\resizebox{\linewidth}{!}{%
\begin{tabular}{@{}crccccccccc@{}}
\toprule
 & \multicolumn{1}{c}{\textbf{}} & \multicolumn{9}{c}{\textbf{Classifier}} \\ \cmidrule(l){3-11} 
\multicolumn{2}{l}{\textbf{Data set}} & \textbf{NB} & \textbf{LR} & \textbf{kNN} & \textbf{DT} & \textbf{RF} & \textbf{GBM} & \textbf{LDA} & \textbf{SVM} & \textbf{MLP} \\ \midrule
 & \textbf{O0} &  76.09 &   74.59 & 78.30   & 85.55 & \textbf{89.04}  & 89.00   &  87.10  &  81.67 &  80.28  \\
\multirow{2}{*}{\textbf{A}} & \textbf{Os} &   72.84 &  73.02  & 79.61  & 86.47 & 83.40  &  \textbf{86.91} & 80.35  & 77.53  &  77.54   \\
 & \textbf{O3} &  72.88   &  74.94 & 83.99 & 85.10 & \textbf{87.42}  & 86.80 &  76.75 & 74.32 &  75.02 \\
 & \textbf{Mixed} &  66.87 & 66.23  & 83.60   & 82.16 &   \textbf{83.81}&    75.68          &   70.07           &     68.15         &    67.22  \\ \midrule
& \textbf{O0} &  82.83 &  83.46  & 82.02  & 90.87 & \textbf{93.34}  & 91.99 & 86.57  & 81.05 &  83.30 \\
\multirow{2}{*}{\textbf{B}} & \textbf{Os} & 80.19 & 78.67 & 79.44 & 87.23 & \textbf{90.04} & 90.03 & 75.11 & 72.84 & 75.56 \\
 & \textbf{O3} &   81.01 &  76.48  &   \textbf{88.29} & 86.32 & 86.37 &  87.41 &  79.85 & 74.32  & 80.20 \\
 & \textbf{Mixed} & 76.74 & 74.02 & 69.81 & 84.59 & \textbf{86.22} & 86.21 & 70.03 & 71.40 & 70.90 \\ \bottomrule
\end{tabular}
}
\end{table}

The data collection and training processes were orchestrated by a Python script using Scikit-Learn~\cite{scikit-learn} on a workstation with an Intel i7-6700k CPU (quad-core at 4.0GHz) and 16GB RM, taking approximately 14 hours (data collection) and 30 hours (training). Classification results for each platform and GCC compilation setting are presented in Tables~\ref{tab:x86_classification_results} (X86-64), \ref{tab:arm_classification_results} (ARM), and \ref{tab:riscv_classification_results} (RISC-V). In general, functions could be classified from HPC values with considerable effectiveness, albeit with notable differences with respect to the architecture and privilege mode.  Worst-case performances occurred where only unprivileged counters were used with the mixed GCC data sets, i.e.\ 48.29\% (X86-64; one HPC) and 83.81\% (RISC-V; three HPCs). Mixed compilation parameters generally correlated with a significant accuracy degradation of 2--16\% depending on the architecture and privilege mode.  In contrast, the best cases corresponded to scenarios where all HPCs were used for specific compilation settings: 97.33\%--99.70\% (X86-64), 90.68\%--96.81\% (ARM), and 86.22--93.34\% (RISC-V). We can tentatively conclude that using more HPCs confers greater discriminitive power during classification, the availability of which are maximised during privileged execution. Further, tree-based models and ensembles tended to perform best out of all evaluated classifiers (18/20) with RF classifiers the best-performing (11/20). 

Confusion matrices were produced for examining class-level performance weaknesses using the mixed GCC data sets and the best-performing classifiers (see Fig.~\ref{fig:confusion_matrix_x86} for X86-64; Appendix~\ref{sec:other_confusion_matrices} for ARM and RISC-V). Confusion is observed in functions with structural similarities when fewer counters are used (Fig.~\ref{fig:classify_x86_unpriv}). For instance, the encryption and decryption functions for ChaCha20-Poly1305, DES, 3DES, and AES in certain modes of operation (e.g.\ CTR and CBC), and RSA encryption, decryption, signature and verification. These errors resolved when more HPCs were available, where greatly reduced classification accuracy was observed.

\subsection{Feature Importance}

An important consideration is the contribution of each HPC feature to the classification process. If only a small set of HPCs is needed to train high accuracy models, then certain implementation difficulties can be avoided.  Minimising the number of hardware counters is desirable for two reasons: 
\begin{enumerate}
    \item \emph{Reproducibility}: Using all of the counters on a given architecture risks depending on redundant HPCs that are unavailable on other architectures, particularly older and constrained platforms. ARM and RISC-V micro-controller and IoT SoCs, for instance, contain fewer measurable events relative to workstation- and server-grade X86-64 CPUs~\cite{riscv:privileged,intel:sys_guide,arm:cortexa53_manual}.
    \item \emph{Performance}: Removing redundant features, or those with low predictive power, can offer training and classification performance benefits due to the curse of dimensionality. (Dimensionality reduction methods have been applied in HPC malware classification literature, e.g.\ principal component analysis~\cite{singh2017detection,zhou2018hardware}, but this does not directly reduce the number of HPCs used at source).
\end{enumerate}

We note that HPC-based research tends to rely on complex, non-linear models, such as random forests and gradient boosting machines, to model high-dimensional data~\cite{zhou2018hardware} (also observed in \S\ref{sec:classification_results}). Unfortunately, these models have decision processes that are inherently difficult to interpret for determining the most effective features, prompting the development of model explanation methods~\cite{shapley:lundberg,shrikumar2017learning}.

\subsubsection{HPC Correlation Analysis} 
\label{sec:correlation_analysis}
As a first step, we examined the correlations between HPC measurements on each platform. It is intuitive that certain features may exhibit significant collinearities (a historically successful method of identifying redundant features~\cite{hall1999correlation}). For example, the total number of load/store instructions (\texttt{LST\_INS}) is directly related to the number of load instructions (\texttt{LD\_INS}). Likewise, processes that frequently access temporally and spatially dislocated data will cause cache misses and, thus, more cache data writes. Fig.~\ref{fig:hpc_corrs} presents pairwise correlation matrices using the correlation coefficients of HPC tuples. 

The results show strong correlations between many HPC pairs. On X86-64, the cycle and time-stamp counters (\texttt{TOT\_CYC} and \texttt{RDTSCP}) are $>$0.95 correlated with the total instruction (\texttt{TOT\_}, \texttt{LD\_}, \texttt{SR\_}, \texttt{BR\_} and \texttt{LST\_INS}) and branch counters (\texttt{BR\_TKN}, \texttt{\_NTK}, and \texttt{\_PRC}). A similar pattern was found for ARM. This is intuitively unsurprising considering that larger, longer-running functions will likely contain more run-time branches and memory accesses. Strong correlations are also seen between cache misses in lower cache hierarchy elements and accesses in higher ones; for example, L2 data accesses and L1 data misses (\texttt{L2\_DCA} and \texttt{L1\_DCM}, 0.96; X86-64). On ARM, L1 \emph{instruction} cache misses are strongly correlated (0.98) with L2 \emph{data} accesses (\texttt{L1\_ICM} vs.\ \texttt{L2\_DCA}). In the absence of a dedicated L2 instruction cache~\cite{arm:cortexa53_manual}, this indicates that the L2 data cache is used as a \emph{de facto} instruction cache, echoing existing work on using HPCs to uncover latent SoC properties~\cite{maurice2015reverse,helm2020reliable}.

It is also seen that cache misses in last-level caches (LLC)---L3 for X86-64 and L2 for ARM (\texttt{L3\_TCM} and \texttt{L2\_DCM} respectively)---have no correlations with other HPCs. PAPI monitors \emph{CPU-level} events, and the LLC represents the final unit before external memories are accessed. Similar patterns exist for TLB data misses (\texttt{TLB\_DM}) on ARM and instruction misses (\texttt{TLB\_IM}) on X86-64 and ARM; and the use of vector and floating point operations on X86-64 (\texttt{SP\_OPS}, \texttt{DP\_OPS}, \texttt{VEC\_SP}, \texttt{VEC\_DP}), which only have correlations with each other. On RISC-V, strong correlations ($>$0.97) were observed between M-mode counters of their U-mode counterparts, e.g.\ \texttt{RDINSTRET} vs.\ \texttt{MINSTRET}. This is interesting from a security perspective: unprivileged processes can use HPCs with potentially the same power as privileged processes. It also provides insight into why classification results for privileged HPCs were only marginally different ($\sim$4\%) from unprivileged HPCs in \S\ref{sec:classification_results}.

\begin{figure}
    \centering
    \begin{subfigure}{\linewidth}
        \centering
        \includegraphics[width=1.04\linewidth]{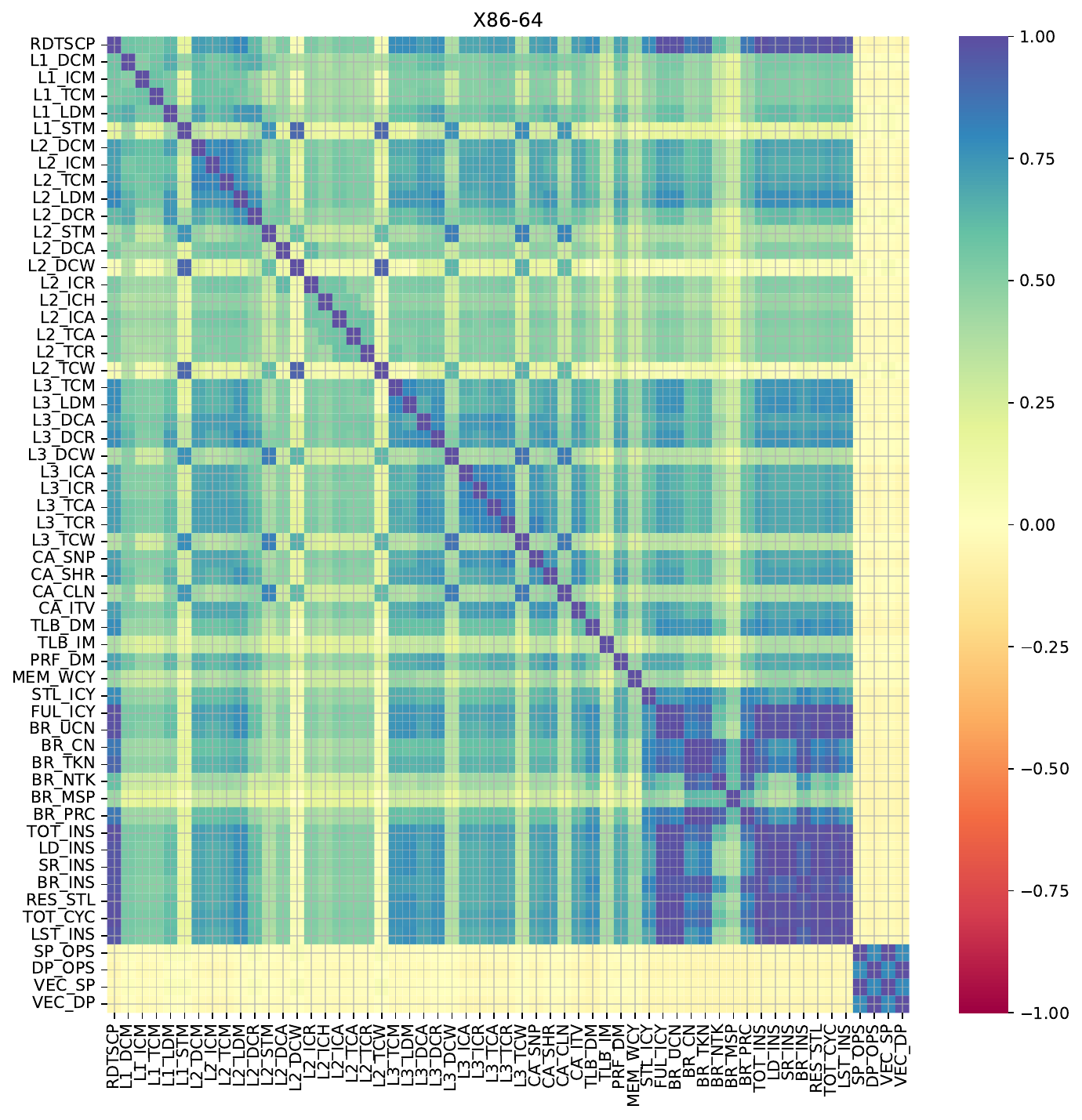}
    \caption{X86-64 HPCs.}
        \label{fig:x86_hpcs}
    \end{subfigure}
    \begin{subfigure}[b]{0.585\linewidth}
        \centering
        \includegraphics[width=\textwidth]{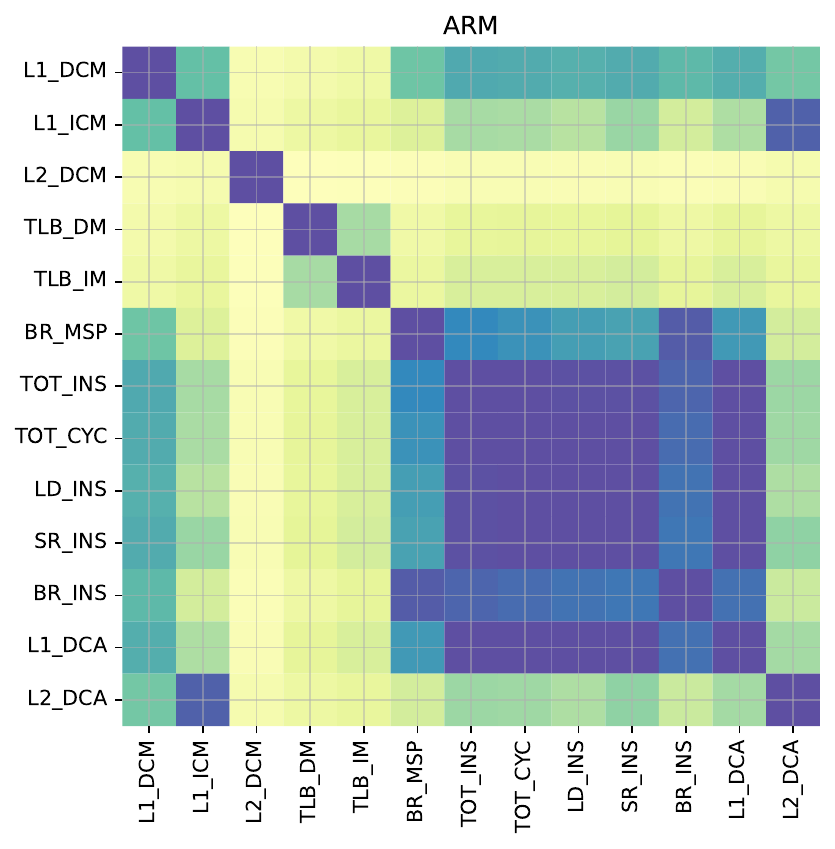}
        \caption{ARM HPCs.}
        \label{fig:arm_hpcs}
    \end{subfigure}
    \begin{subfigure}[b]{0.405\linewidth}
         \centering
         \includegraphics[width=\textwidth]{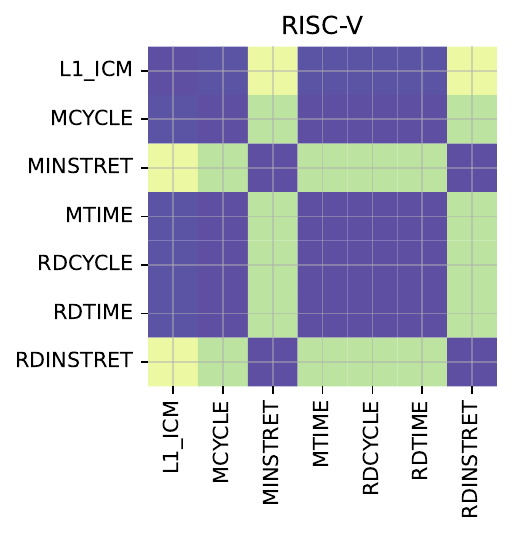}
         \caption{RISC-V HPCs.}
         \label{fig:riscv_hpcs}
    \end{subfigure}
    \caption{Correlation matrices for each test-bed device.}
    \label{fig:hpc_corrs}
\end{figure}

\subsubsection{Shapley Additive Explanations}
\label{sec:shap}

 While useful for understanding multicollinearity, correlations do not show the extent to which particular HPCs contribute to classification decisions, thus requiring model inspection methods to understand their relative importance.  To this end, we employed SHAP (SHapley Additive exPlanations), a unified framework for interpreting complex model predictions, which has been applied to Android malware classification~\cite{fan2020can} and intrusion detection~\cite{wang2020explainable}. SHAP uses a co-operative game-theoretic approach for assigning feature importances of a machine learning model prediction function, $f$, using Shapley values~\cite{shapley:lundberg}. SHAP explains $f$'s decisions as the sum of results, $\phi_i \in \mathbb{R}$, of feature subsets being included in a conditional expectation, $E[f(x) | x_{S}]$ ($S$ being a subset of model features and $x$ the feature vector of the instance to be explained). SHAP scores combine conditional expectations with the Shapley value of a feature value, $\phi_i$---corresponding to its contribution to the payout (prediction)---which are calculated using Eq.~\ref{eq:shapley}.

\begin{equation}
    \phi_i = \sum_{S \subseteq N \backslash \{i\}} \frac{| S |! (M-|S|-1)!}{M!} [f_x (S \cup \{i\} ) - f_x (S)]
    \label{eq:shapley}
\end{equation}

Where $M$ is the number of features and $N$ is the set of all input features.  The average absolute Shapley values per feature are computed across the entire data set and rank-ordered to find the global feature importances. SHAP values are more consistent with human intuition than alternative approaches, like LIME~\cite{ribeiro2016should} and DeepLIFT~\cite{shrikumar2017learning}; are model-agnostic; and are not liable to the properties of HPC measurements. (Other techniques, e.g.\ mean decrease in Gini impurity for tree models, can be misleading with high cardinality or continuous features~\cite{strobl2007bias}---an inherent property of HPCs). 

We computed and rank-ordered the SHAP values for each platform HPC using the best-performing classifiers from \S\ref{sec:classification_results} under the mixed GCC data sets (Fig.~\ref{fig:hpc_corrs}). Certain HPCs had a significant impact on classification decisions across all platforms. Branch counters formed the top two ARM HPCs (\texttt{BR\_MSP}, \texttt{BR\_INS}) and six of the top 10 X86-64 counters (\texttt{BR\_PRC}, \texttt{BR\_CN}, \texttt{BR\_NTK}, \texttt{BR\_TKN}, \texttt{BR\_INS}, \texttt{BR\_UCN}). The RISC-V platform could not measure branch events.  Instruction counters also ranked highly, representing the top two RISC-V HPCs (\texttt{RDINSTRET}, \texttt{MINSTRET}), three of the top five ARM HPCs (\texttt{BR\_INS}, \texttt{SR\_INS}, \texttt{TOT\_INS}), and three of the top 10 X86-64 HPCs (\texttt{TOT\_INS}, \texttt{BR\_INS}, \texttt{SR\_INS}). We observe that cache events ranked relatively poorly, particularly TLB data and instruction misses (X86-64 and ARM), and LLC events (ARM L2 and X86-64 L3).

\begin{figure}
\centering
    \begin{subfigure}{0.83\linewidth}
        \centering
        \includegraphics[width=\textwidth]{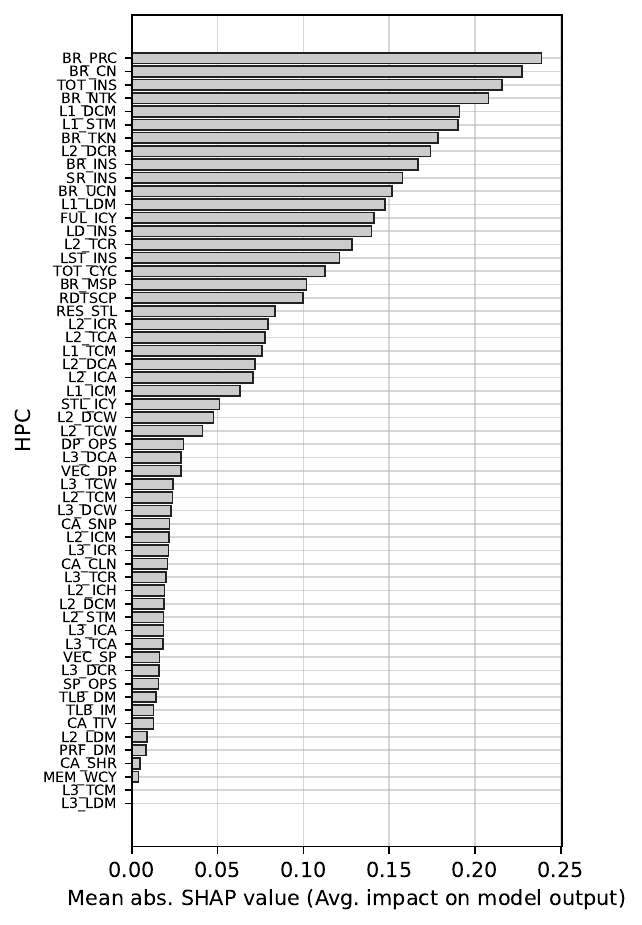}
        \caption{X86-64 HPCs.}
        \label{fig:x86_hpcs}
    \end{subfigure}
    \begin{subfigure}{0.495\linewidth}
        \centering
        \includegraphics[width=1.05\textwidth]{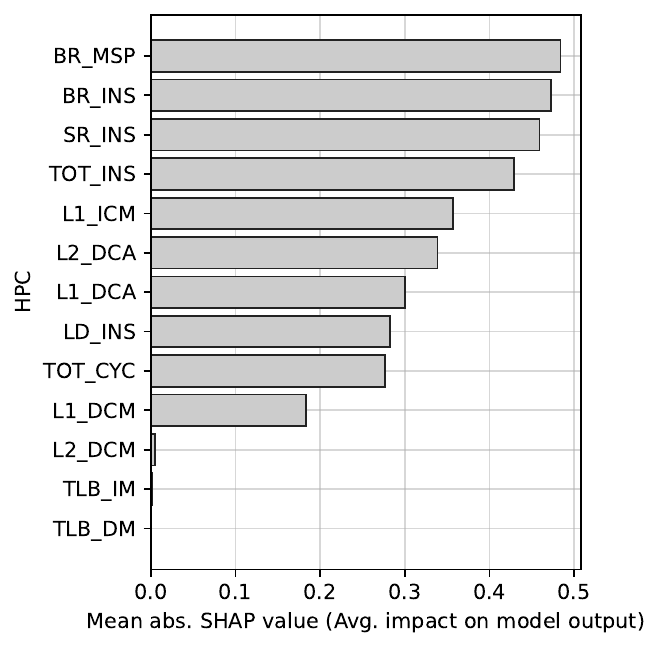}
        \caption{ARM HPCs.}
        \label{fig:arm_hpcs}
    \end{subfigure}
    \begin{subfigure}{0.495\linewidth}
         \centering
         \includegraphics[width=1.05\textwidth]{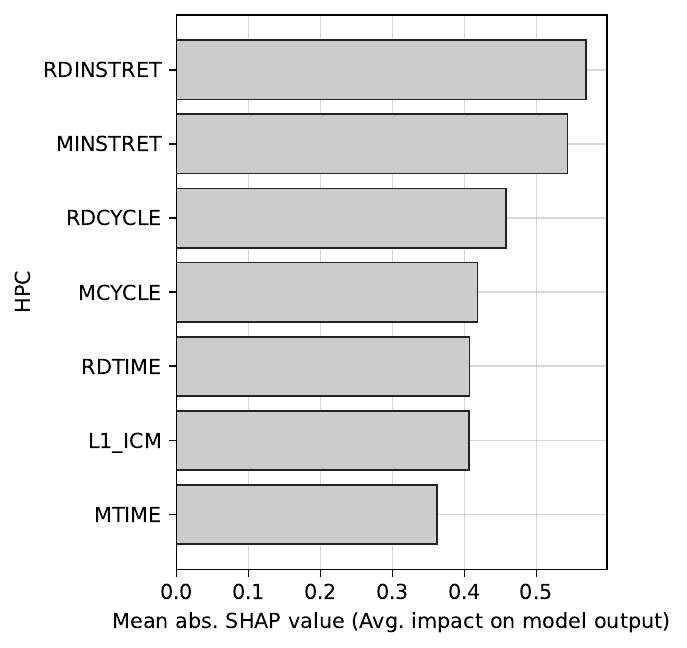}
         \caption{RISC-V HPCs.}
         \label{fig:riscv_hpcs}
    \end{subfigure}
    \caption{Rank-ordered mean absolute SHAP values.}
    \label{fig:hpc_corrs}
\end{figure}

\subsubsection{Feature Elimination}

SHAP values gauge the relative contribution of features, but do not directly determine an effective minimal set of HPCs required for strong model performance.  Therefore, we investigated how model accuracy fluctuated by systematically including particular hardware counters. Here, the best performing models from \S\ref{sec:classification_results} were retrained using the top $N$ features from the SHAP analysis under the same classifier hyper-parameters. Values of $N$ were evaluated in the range 1--10 and compared with using all available HPCs. 

\begin{figure}[t]
    \centering
    \includegraphics[width=\linewidth]{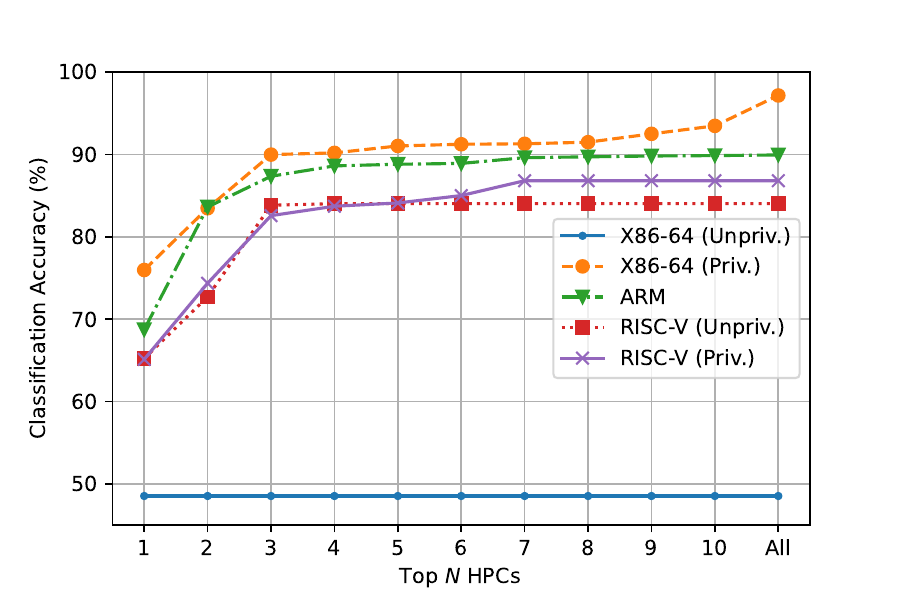}
    \caption{Accuracy using the top $N$ SHAP-valued HPCs.}
    \label{fig:retraining_accuracy}
\end{figure}

The results are shown in Fig.~\ref{fig:retraining_accuracy} for each platform using HPCs accessible in unprivileged and privileged mode.  Evidently, using all HPCs bestows \emph{some} accuracy benefits, but only a subset were necessary for achieving close results. Using all 49 X86-64 counters, for example, delivered a $\sim$5\% improvement over using the top 10 from Fig.~\ref{fig:x86_hpcs}.  The top three HPCs achieved $<$9\% of the final accuracy on X86-64 (\texttt{BR\_PRC}, \texttt{BR\_CN}, \texttt{TOT\_INS}) while only two ARM counters were needed to achieve this (\texttt{BR\_MSP}, \texttt{BR\_INS}). On RISC-V, the accuracy benefits decreased to $\sim$5\% using three counters (\texttt{RDINSTRET}, \texttt{MISNTRET}, \texttt{RDCYCLE}) versus all seven.

Our results generally indicate that a large range of black-box functions can be classified effectively based on their HPC values---up to 97.33\%---across three major CPU platforms. This varies significantly according to privilege mode and counter availability on our chosen platforms.  In the coming sections, we explore applications of this generic approach in the domains of vulnerability detection within OpenSSL and function recognition within TEE applications.

%% file: sections/openssl-vulnerabilities.tex
\section{Identifying Known Security Vulnerabilities in OpenSSL}
\label{sec:openssl_vulns}

Cryptographic libraries are often deployed as static or shared libraries in end-user applications. OpenSSL, for instance, is deployed as \texttt{libssl} and \texttt{libcrypto}, with the latter implementing dependent cryptographic algorithms which may be used in isolation. By default, both libraries are compiled and named using major and minor version numbers, e.g.\ 1.0, 1.1, and 3.0. Incremental sub-versions are also released after remedying known security vulnerabilities, known as \emph{lettered} releases. 

In this use case, we examine the extent to which vulnerable \texttt{libcrypto} lettered versions can be detected using HPC measurements from invoking affected cryptographic functions.  Specifically, we examine the extent to which OpenSSL vulnerabilities can be identified using HPC differences in calls to patched and unpatched \texttt{libcrypto} functions.  This can be used as an exploratory technique for isolating useful attack vectors in the dependencies of application binaries. Broadly, the attacker is assumed to possess the ability to: \Circled{1} instrument the target binary to measure function calls to the dependent library; or \Circled{2} compile and link his/her own measurement harness to invoke functions in the dependency.

\subsection{Methodology}

Our developed approach follows three phases:
\begin{enumerate}
    \item \emph{Preliminary vulnerability identification.} We comprehensively examined the OpenSSL vulnerability disclosure announcements\footnote{\url{https://www.openssl.org/news/vulnerabilities.html}} to identify vulnerabilities that cause internal micro-architectural state changes, but whose effects are not immediately observable (e.g.\ does not induce a crash, infinite loop, or returns certain function values).  Vulnerable/patched functions were located using NIST's National Vulnerability Database (NVD), which aggregates technical write-ups, third-party advisories, and, importantly, commit-level patch details for CVEs.  In total, we successfully identified six vulnerabilities, which may be used for DSA, RSA, and ECDSA private key recovery. The descriptions, CVE numbers, and commit IDs are given in Appendix~\ref{sec:cve_description}. 
    \item \emph{Collecting labelled data samples}.  For each vulnerability, we collected HPC measurements of 10,000 executions using all available counters (privileged and unprivileged). We measured offending \texttt{libcrypto} functions of the same major and minor version, but from different \emph{lettered} versions before and after the patch was implemented. For example, for a function patched in \texttt{v1.1.0f}, measurements were taken from the preceding \texttt{v1.1.0a-e} (unpatched) and successive versions \texttt{v1.1.0f-n} (patched). This required compiling and linking our PAPI measurement harness against multiple individual \texttt{libcrypto} lettered versions. In contrast to \S\ref{sec:algorithm_identification}, which considered functon recognition as a multi-class problem, vulnerability detection is treat as a \emph{binary} problem, where measurement vectors were assigned labels in the range $[0,1] \in \mathbb{N}$ (0 = patched, 1 = unpatched).
    \item \emph{Model selection and evaluation}. Following the same method as \S\ref{sec:algorithm_identification_methodology}, multiple models were trained using an 80:20 training-test set ratio, 10-fold cross-validation, and exhaustive grid search for hyper-parameter optimisation. In addition to classification accuracy, precision, recall, and F1-score metrics were employed for evaluating binary classification performance. This is important when considering the unbalanced nature of vulnerability detection in this context. Measurement data sets of vulnerabilities remedied in early OpenSSL lettered versions, e.g.\ \texttt{1.1.0b}, will contain far fewer `unpatched' labels than those in later versions, thus necessitating evaluation metrics that consider relevance. 
\end{enumerate}

\subsection{Results and Analysis}

We applied the methodology to known vulnerabilities within \texttt{libcrypto} using the same target devices from \S\ref{sec:algorithm_identification}.\footnote{OpenSSL does not support RISC-V at the time of publication.} The classification results are presented in Tables \ref{tab:openssl-results-x86} and \ref{tab:openssl-results-arm} for X86-64 and ARM respectively for each vulnerability. Our approach identified OpenSSL vulnerabilities with high accuracy: 89.5\% in all cases and 93\%+ accuracy in all but one case across both architectures. Likewise, precision (0.892--0.997, X86-64; 0.882--0.978, ARM), recall (0.900--0.998, X86-64; 0.873--0.995, ARM) and F1 scores (0.896--0.998, X86-64; 0.889--0.985, ARM) were consistently high, demonstrating effectiveness when identifying feature vectors corresponding to unpatched instances.

It is noteworthy that some differences exist between each identified CVE, particularly \texttt{CVE-2018-0737}, which has significantly lower accuracy (4--5\%) than the next worst performing CVE. On closer inspection, we noticed a broad correlation between classification performance and relative patch complexity. This is not surprising: security-related patches with significant code additions and/or deletions will cause deterministic effects in HPC measurements. As an elementary example, additional conditional statements will correlate with increased values of branch-related, cycle, and instruction counters. The \texttt{CVE-2018-0737} patch---the worst performing vulnerability---changed only two lines of code (LOC) between lettered OpenSSL versions for setting constant-time operation flags for RSA primes (commit 6939eab03a6e23d2bd2c3f5e34fe1d48e542e787\footnote{\url{https://github.com/openssl/openssl/commit/6939eab03a6e23d2bd2c3f5e34fe1d48e542e787}}).\footnote{LOC is an illustrative proxy of program complexity; a single line may induce complex control flows, with significant effects on HPCs.} Compare this to \texttt{CVE-2018-0734}---detected with the highest accuracy---which made 19 LOC changes, including the declaration of new variables, conditional statements, internal function calls, and more (commit 8abfe72e8c1de1b95f50aa0d9134803b4d00070f\footnote{\url{https://github.com/openssl/openssl/commit/8abfe72e8c1de1b95f50aa0d9134803b4d00070f}}). 

This raises a research question about the \emph{granularity} with which counters can reliably detect arbitrary code changes. The known effects of non-determinism---see \S\ref{sec:hpc_implementation_challenges} and Das et al.~\cite{das2019sok}---and pipeline execution on measurement noise~\cite{arm:cortexa53_manual} provide an undetermined lower bound. Yet, this has not been answered in related literature, which we pose as an obvious gap for future research.  Notwithstanding, our approach shows that HPC measurements of function calls can detect known vulnerabilities using off-line analysis.

\begin{table}
\centering
\caption{OpenSSL CVE identification results (X86-64).}
\label{tab:openssl-results-x86}
\begin{threeparttable}
\begin{tabular}{@{}rccccc@{}}
\toprule
\textbf{CVE ID} & \textbf{Operation} & \textbf{Pr.} & \textbf{Re.} & \textbf{F1} & \textbf{Acc.} \\ \midrule
\texttt{CVE-2018-5407} & ECC scalar mul.\ & 0.915 & 0.977 & 0.945 & 94.67 \\
\texttt{CVE-2018-0734} & DSA sign & 0.997 & 0.998 & 0.998 & 99.83 \\
\texttt{CVE-2018-0735} & ECDSA sign & 0.974 & 0.950 & 0.962 & 97.50 \\
\texttt{CVE-2018-0737} & RSA key gen.\ & 0.892  & 0.900 & 0.896 & 90.25 \\
\texttt{CVE-2016-2178} & DSA sign & 0.985 & 0.978 & 0.981 & 98.75  \\
\texttt{CVE-2016-0702} & RSA decryption & 0.940 & 0.938 & 0.939 & 95.92  \\ \bottomrule
\end{tabular}
\begin{tablenotes}
\item Pr.: Precision, Re.: Recall, F1: F1-score, Acc.: Accuracy (in \%).
\end{tablenotes}
\end{threeparttable}
\end{table}
\begin{table}
\centering
\caption{OpenSSL CVE identification results (ARM).}
\label{tab:openssl-results-arm}
\begin{tabular}{@{}rccccc@{}}
\toprule
\textbf{CVE ID} & \textbf{Operation} & \textbf{Pr.} & \textbf{Re.} & \textbf{F1} & \textbf{Acc.} \\ \midrule
\texttt{CVE-2018-5407} & ECC scalar mul.\ & 0.891 & 0.975 & 0.931 & 93.25 \\
\texttt{CVE-2018-0734} & DSA sign & 0.976 & 0.995 & 0.985 & 99.00  \\
\texttt{CVE-2018-0735} & ECDSA sign & 0.978 & 0.910 & 0.943 & 96.33  \\
\texttt{CVE-2018-0737} & RSA key gen.\ & 0.882 & 0.896 & 0.889 & 89.58  \\
\texttt{CVE-2016-2178} & DSA sign & 0.954 & 0.938 & 0.946 & 96.42  \\
\texttt{CVE-2016-0702} & RSA decryption & 0.939 & 0.873 & 0.904 & 93.83  \\ \bottomrule
\end{tabular}
\end{table}

%% file: sections/tee.tex
\section{Recognising Cryptographic Algorithms in a Trusted Execution Environment (TEE)}
\label{sec:tee_reverse}

The previous section showed how some security vulnerabilities can be detected within compiled OpenSSL (\texttt{libcrypto}) libraries. The general approach is extendable to recognising functions executing within trusted execution environment (TEE) applications. TEE applications typically expose high-level APIs for enabling untrusted applications to interact with secure world services~\cite{trustedfirmware,arm:cortexa53_manual,gp:tee}. In this section, we examine an application of our approach to recognise standardised cryptographic functions within ARM TrustZone TEE applications from a malicious, non-secure world process.

\subsection{ARM TrustZone and the ARM PMU}

 ARM TrustZone partitions platform execution into `secure' (SW) and `non-secure' (NS) worlds, with the aim  of protecting SW service services from kernel-level, non-secure world software attacks.  SW execution is isolated by setting the NS-bit by the secure monitor at the highest ARM exception (privilege) level (EL3), which is added to cache tags and propagated through system-on-chip bus transactions, e.g.\ for accessing sensitive peripheral controllers.  Interactions between the normal and secure worlds are conducting using ARM secure monitor calls (SMC) for entering secure monitor mode. NS world applications invoke TA functions using a pre-defined interfaces specified by the TA developer, as standardised by the GlobalPlatform Client API~\cite{gp:tee}.  Notably, TEE TAs are usually provisioned in encrypted form, which are subsequently loaded from flash memory during the device's secure boot sequence using a firmware-bound key. This occurs \emph{before} loading any untrusted world binaries, rendering direct inspection of TA binaries tremendously difficult, even from privileged NS world execution~\cite{shepherd:physical_fias_scas,trustedfirmware}.

Recall from \S\ref{sec:arm_counters} that the ARM PMU manages performance events on ARM Cortex-A platforms. Ideally, PMU interrupt events should be suppressed during secure world execution to prevent sensitive micro-architectural state changes being measurable from malicious non-secure world processes. However, enabling SW PMU events is commonly used in pre-release testing environments for TEE debugging and optimisation (a non-invasive method under the ARM debugging architecture~\cite{ning2019understanding}). Whether or not SW PMU events are enabled can be determined by querying the non-invasive and secure non-invasive flags (\texttt{NIDEN} and \texttt{SPNIDEN}) of the ARM \texttt{DBGAUTHSTATUS} debug register. If \texttt{NIDEN} or \texttt{SPNIDEN} are set, then PMU events are counted in the non-secure and secure worlds respectively. 

Importantly, Spisak~\cite{spisak2016hardware} showed that \texttt{SPNIDEN} was still enabled in some consumer devices following release, including the Amazon Fire HD 7" tablet and Huawei Ascend P7. This was followed by Ning and Zhang~\cite{ning2019understanding} who examined 11 mobile, IoT and ARM server platforms, showing that only three devices had correctly unset \texttt{SPNIDEN} prior to consumer release. Vulnerable devices included the Huawei Mate 7, Raspberry Pi 3B+, Xiaomi Redmi 6, and the Scaleway ARM C1 Server. The use of insecurely configured ARM PMU events during secure world execution has been exploited on consumer devices for building rootkits using PMU interrupts~\cite{spisak2016hardware}, and cross-world covert channels on an undisclosed Samsung Tizen TV (ARM Cortex-A17, ARMv7) and HiKey board (ARM Cortex-A53, ARMv8)~\cite{cho2018prime}.

\subsection{Methodology}

Using the knowledge that consumer devices may fail to suppress SW PMU interrupts, we developed a test-bed for investigating the extent to which cryptographic algorithms within secure world TAs can be identified from non-secure world processes. OP-TEE\footnote{\url{https://op-tee.org}} was leveraged to this end---an open-source, GlobalPlatform-compliant TEE reference implementation---on our Raspberry Pi 3B+. Two applications were developed, which are illustrated in Fig.~\ref{fig:ta-spy-flow}:

\begin{figure}
    \centering
    \includegraphics[width=\linewidth]{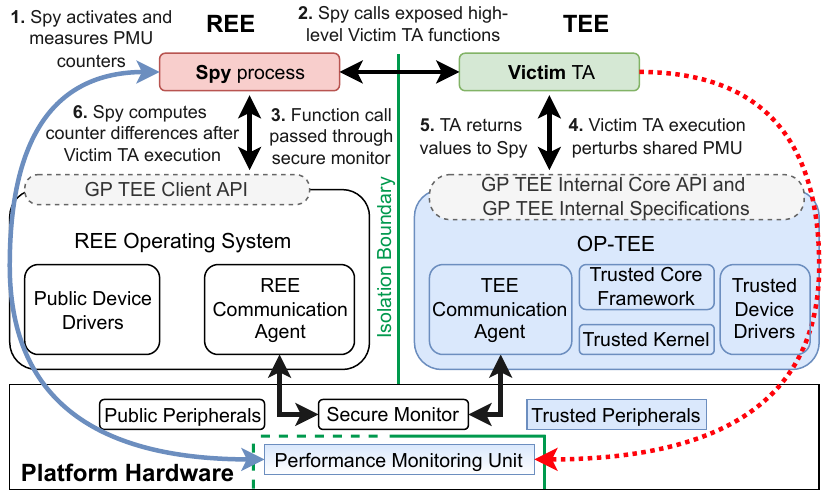}
    \caption{TEE side-channel set-up using a shared PMU.}
    \label{fig:ta-spy-flow}
\end{figure}

\begin{itemize}
    \item \emph{Spy} (non-secure world). A privileged application that uses the GlobalPlatform Client API to invoke TEE functions in the \emph{Victim}. The same PAPI test harness used in \S\ref{sec:algorithm_identification} and \S\ref{sec:openssl_vulns} measured HPC values immediately before and after TA command was invoked, i.e.\ \texttt{TEEC\_InvokeCommand()} in GlobalPlatform Client API nomenclature. The \emph{Spy} may be unprivileged if the PMU is configured to enable measurements from EL0.
    \item \emph{Victim} (secure world TA). An emulation of a TEE-based key management application that exposes four high-level functions for \Circled{1} signing (\texttt{TA\_SIGN}), \Circled{2} verifying (\texttt{TA\_VERIFY}), \Circled{3} encrypting (\texttt{TA\_ENCRYPT}), and \Circled{4} decrypting (\texttt{TA\_DECRYPT}) inputs provided by the \emph{Spy} on demand. These utilised fresh TEE-bound keys which were generated at random on a per-session basis.\footnote{Under the GlobalPlatform TEE architecture, a non-secure client application invokes one or more \emph{commands}, e.g.\ signing or decryption, of a target TA within a single \emph{session}.}
\end{itemize}

The aim, similar to \S\ref{sec:algorithm_identification}, is to identify the precise algorithm used by the \emph{Victim} from PMU measurements before and after its invocation by the \emph{Spy}. The \emph{Spy} is given only the aforementioned high-level functions for signing, verification, encryption, and decryption. Secure world cryptographic functions are implemented in OP-TEE using LibTomCrypt whose functions are wrapped by the GlobalPlatform Internal Core API~\cite{gp:tee}. The \emph{Victim} was developed to call an extensive range of GlobalPlatform Internal Core API functions implemented by the OP-TEE OS core. For calculating performance results, the \emph{Victim} TA also accepted a given Internal Core API algorithm ID, which was used for labelling HPC vectors. The full list of analysed cryptographic algorithms is given in Table~\ref{tab:optee-algs}, covering all available modes of operation and padding schemes where applicable. 

Using our previously developed PAPI test harness, HPC measurements of 1,000 invocations of each algorithm were collected from the non-secure world using all available HPCs on our platform. This was repeated for 100 sessions, with the data collected afterwards for off-line analysis (100,000 vectors per algorithm; 3.4M in total). The measurement vectors were labelled with respect to each GlobalPlatform Internal Core algorithm identifier, i.e.\ with the label $[0,34) \in \mathbb{N}$. Like \S\ref{sec:algorithm_identification}, key sizes were not fixed; a random key size was set during the algorithm's run-time allocation prior to its execution. It is important to note that OP-TEE TAs are cross-compiled using a GCC-based toolchain, rendering them vulnerable to the compilation biases (\S\ref{sec:gcc}). To address this, we evaluated the GCC optimisation flags from \S\ref{sec:algorithm_identification} in order to assess the effects of different optimisation levels on classification performance.

\begin{table}
\centering
\caption{\emph{Victim} TA algorithms and key sizes (bits).}
\label{tab:optee-algs}
\resizebox{\columnwidth}{!}{%
\begin{tabular}{@{}r|l|c@{}}
\toprule
\textbf{Method} & \textbf{GlobalPlatform Internal Core API ID} & \textbf{Key Sizes}\\ \midrule
\multicolumn{3}{c}{\textbf{TA\_SIGN and TA\_VERIFY}} \\\midrule
\multirow{3}{*}{DSA} & TEE\_ALG\_DSA\_SHA1 & 512, 1024 \\
 & TEE\_ALG\_DSA\_SHA224 & 2048 \\
 & TEE\_ALG\_DSA\_SHA256 & 2048, 3072 \\\midrule
\multirow{5}{*}{ECDSA} & TEE\_ALG\_ECDSA\_P192 & 192 \\
 & TEE\_ALG\_ECDSA\_P224 & 224 \\
 & TEE\_ALG\_ECDSA\_P256 & 256\\
 & TEE\_ALG\_ECDSA\_P384 & 384 \\
 & TEE\_ALG\_ECDSA\_P521 & 512\\\midrule
\multirow{11}{*}{RSA} & TEE\_ALG\_RSASSA\_PKCS1\_V1\_5\_MD5 & \\
 & TEE\_ALG\_RSASSA\_PKCS1\_V1\_5\_SHA1   & \\
 & TEE\_ALG\_RSASSA\_PKCS1\_V1\_5\_SHA224 & \\
 & TEE\_ALG\_RSASSA\_PKCS1\_V1\_5\_SHA256 & \\
 & TEE\_ALG\_RSASSA\_PKCS1\_V1\_5\_SHA384 & \\
 & TEE\_ALG\_RSASSA\_PKCS1\_V1\_5\_SHA512 & 1024, 2048 \\
 & TEE\_ALG\_RSASSA\_PKCS1\_PSS\_MGF1\_SHA1 & 3072, 4096\\
 & TEE\_ALG\_RSASSA\_PKCS1\_PSS\_MGF1\_SHA224 & \\
 & TEE\_ALG\_RSASSA\_PKCS1\_PSS\_MGF1\_SHA256 & \\
 & TEE\_ALG\_RSASSA\_PKCS1\_PSS\_MGF1\_SHA384 & \\
 & TEE\_ALG\_RSASSA\_PKCS1\_PSS\_MGF1\_SHA512 & \\\midrule
\multicolumn{2}{c}{\textbf{TA\_ENCRYPT and TA\_DECRYPT}} \\\midrule
\multirow{5}{*}{AES} & TEE\_ALG\_AES\_CBC\_NOPAD & \\
 & TEE\_ALG\_AES\_CCM & \\
 & TEE\_ALG\_AES\_CTR & 128, 196 \\
 & TEE\_ALG\_AES\_ECB\_NOPAD & 256 \\
 & TEE\_ALG\_AES\_GCM & \\\midrule
\multirow{2}{*}{DES} & TEE\_ALG\_DES\_ECB\_NOPAD  & \multirow{2}{*}{64} \\
 & TEE\_ALG\_DES\_CBC\_NOPAD & \\\midrule
\multirow{2}{*}{3DES} & TEE\_ALG\_DES3\_ECB\_NOPAD & \multirow{2}{*}{128, 192}\\
 & TEE\_ALG\_DES3\_CBC\_NOPAD & \\\midrule
\multirow{6}{*}{RSA} & TEE\_ALG\_RSAES\_PKCS1\_V1\_5 & \\
 & TEE\_ALG\_RSAES\_PKCS1\_OAEP\_MGF1\_SHA1 & \\
 & TEE\_ALG\_RSAES\_PKCS1\_OAEP\_MGF1\_SHA224 & 1024, 2048 \\
 & TEE\_ALG\_RSAES\_PKCS1\_OAEP\_MGF1\_SHA256 & 3072, 4096 \\
 & TEE\_ALG\_RSAES\_PKCS1\_OAEP\_MGF1\_SHA384 & \\
 & TEE\_ALG\_RSAES\_PKCS1\_OAEP\_MGF1\_SHA512 & \\ \bottomrule
\end{tabular}
}
\end{table}

\subsection{Results}

After retrieving the data files from the test platform, the same procedure was followed as in the previous sections. The data file was divided into training and test sets using an 80:20 ratio before applying exhaustive grid search with 10-fold cross-validation to select the best performing classifier. The final accuracy was calculated using the performance of the best performing cross-validation classifier on the aforementioned test set. Results of this analysis are given in Table~\ref{tab:ta_classification_accuracy}. The results generally reflect those in \S\ref{sec:algorithm_identification} and \S\ref{sec:openssl_vulns}: algorithm recognition can be achieved with high accuracy using PMU values perturbed by a secure world TA that is measured by a non-secure world application. In the best case, 95.50\% classification accuracy (DT) was achieved for the mixed GCC data-set, increasing slightly to best cases of 96.13\% (RF), 96.02\% (GBM), and 97.45\% (GBM) for the O0, Os, and O3 data sets respectively.

 \begin{table}
\centering
\caption{Classification accuracy using \emph{Spy} (non-secure world) HPC feature vectors from \emph{Victim} (TEE) execution.}
\label{tab:ta_classification_accuracy}
\resizebox{\linewidth}{!}{%
\begin{tabular}{@{}crccccccccc@{}}
\toprule
 & \multicolumn{1}{c}{\textbf{}} & \multicolumn{9}{c}{\textbf{Classifier}} \\ \cmidrule(l){3-11} 
\multicolumn{2}{r}{\textbf{Dataset}} & \textbf{NB} & \textbf{LR} & \textbf{kNN} & \textbf{DT} & \textbf{RF} & \textbf{GBM} & \textbf{LDA} & \textbf{SVM} & \textbf{MLP} \\ \midrule
 & \textbf{O0} &  83.99 &   90.12 & 88.07   & 94.48  & \textbf{96.13}  & 90.76   &  92.02  &  88.54 &  87.90  \\
 & \textbf{Os} &   84.61 &  87.92  & 84.68  & 90.72 & 94.07  & \textbf{96.02} & 95.51  & 85.81  &  86.75   \\
 & \textbf{O3} &  83.18   &  92.49 & 93.19 & 89.58  & 93.33  & \textbf{97.45}  &  79.66 & 94.35 &  87.71 \\
 & \textbf{Mixed} &  82.89 & 85.27  & 92.83  & \textbf{95.50} &  94.17  &    86.93  &  82.55           &     86.99         &     90.82  \\ \bottomrule
\end{tabular}
}
\end{table}

%% file: sections/evaluation.tex
\section{Evaluation}
\label{sec:evaluation}

This section analyses mitigations, limitations, and challenges of the work presented in this paper.

\subsection{Analysis and Mitigations}

Exploiting PMUs as a side-channel medium has been acknowledged by CPU architecture designers and specification bodies. ARM concede that counters are a potential side channel for leaking confidential information, and issue secure development guidance for preventing TEEs from perturbing HPC measurements during TA execution~\cite{trustedfirmware,arm:cortexa53_manual}. Likewise, the RISC-V Unprivileged specification states that \emph{``Some execution environments might prohibit access to counters to impede timing side-channel attacks''}~\cite{riscv:unprivileged} (Chapter 10, p.\ 59). Despite this, PMUs are still widely accessible with privileged access and are still perturbed by TEE TAs on some consumer devices~\cite{spisak2016hardware,ning2019understanding,cho2018prime}.

If attackers are assumed to possess only user-mode execution, then certain system-level countermeasures can be deployed. On X86-64, the \texttt{CR4.TSD} and \texttt{CR4.PCE} control registers can be set and unset to prevent the reading of time-stamp (\texttt{RDTSC}) and programmable PMU counter values (\texttt{RDPMC}) respectively. On Linux devices, the \texttt{kernel.perf\_event\_paranoid} flag can be set to a non-zero value to prevent user-space processes from accessing PMU values through the \emph{perf} subsystem. Alternatively, as a common but blunt countermeasure, \emph{perf} can be disabled at installation time for providing access to high-resolution CPU events.  It is worth stating that performance counters have many legitimate uses, including application benchmarking and debugging, which would be prevented by this mitigation.

Side-channel resistance has been studied extensively in the context of avoiding time- and data-dependent code. Cache-based timing attacks, in particular, have prompted the development of resistant cryptographic algorithms (e.g.\ bit-sliced AES implementations). In recent years, attacks that leverage side-effects of transient execution have compounded these issues~\cite{li2018online,alam2017performance}. From the results in this work, we emphasise that focussing on time-, cache-based, or branch prediction-based side-channel mitigations are insufficient for protecting against the presented methods. Ideally, library and TEE TA developers must consider a larger range of measurable performance events, such as instruction retirements, TLB hits/misses, load/store counts, pre-fetch misses, and memory writes. While certain \emph{individual} events have been exploited and mitigated, \emph{multiple} events can be leveraged to bypass countermeasures against any particular micro-architectural side-channel approach.

For software-based mitigations, inspiration can be taken from Li and Gaudiot~\cite{li2020challenges} who posed the challenge of HPC-based speculative execution attack detection in the presence of `evasive' adversaries. State-of-the-art accuracy of Spectre detection models declined by 30\%--40\% after introducing instructions that mimicked benign programs. Thus, one countermeasure is to introduce micro-architectural obfuscation by randomly and significantly perturbing the PMU during the execution of sensitive functions. The injection of well-crafted injection was also suggested by Liu et al.~\cite{liu2019adaptive} for countering ARM cache-based side-channels and by Carrelli et al.~\cite{carelli2019performance} against certain timing attacks. However, we such stress that noise must apply to \emph{many} HPC events, not just cache- or timing-based counters. How this can be feasibly and effectively achieved is posed as an open challenge.

Attention is also drawn to countermeasures when deploying Intel Software Guard eXtensions (SGX) enclaves. SGX applications can be built using the `anti side-channel interference' (ASCI) feature, which suppresses interrupts to Intel PMUs upon entry to production enclaves~\cite{intel:sys_guide}, thus preventing their use as a side-channel for inspecting enclave contents. Unless developers explicitly and negligently opt-out of ASCI, then production enclaves are strengthened significantly to the attacks described in this paper. Likewise, we reiterate best-practice guidelines to device manufacturers to prevent PMU events being raised during secure world execution by securely configuring the PMU control and debug registers upon TEE entry.

\subsection{Challenges, Limitations, and Practicability}
\label{sec:general_challenges}

Using HPCs to classify program functions faces some interesting challenges that were considered outside the scope of this work. Firstly, we investigated programs with few levels of intermediate abstraction: using C programs on a bare-metal microcontroller (RISC-V) or with a single host OS (Debian-based Linux; ARM and X86-64). Instrumenting and measuring programs using HPCs faces difficulties in virtualised environments sharing a single set of CPU counters; for instance, within virtual machines (VMs) and containers with OS-level virtualisation (e.g.\ Docker). Programs written in interpreted languages and those with just-in-time (JIT) compilation, e.g.\ Java, also pose known challenges for side-channel analysis~\cite{shepherd:physical_fias_scas}. Further research is required to correctly account for the additional noise from multiple users, processors, garbage collectors, etc.\ on a single PMU.

Secondly, a significant number of possible implementations variations may be encountered in reality on an arbitrary platform. The variations were extensive but not exhaustive; for example, the use of hardware implementations; different RNG sources; expanded cryptographic libraries, e.g.\ Crypto++ and Bouncy Castle; and alternative TEE implementations were not evaluated. The scope of this work was not to provide an comprehensive analysis of these possibilities. Rather, we aimed to provide the first investigation into using HPCs for function recognition, while providing a detailed understanding of HPC correlations, relative classification contributions, and the applications to vulnerability detection and TA analysis.  We also wish to briefly note that the recognition of \emph{arbitrary} functions cannot be practically achieved considering that it may not terminate, reflecting the halting problem. For constrained cases, like recognising encryption functions from a set of potential library candidates, the data acquisition and classification stages remained feasible in nature: $<$3 minutes per feature vector in the worst cases of unoptimised RSA encryptions and DH key exchanges on our RISC-V microcontroller.

Thirdly, in a practical scenario, the generic approaches in \S\ref{sec:algorithm_identification} and \S\ref{sec:tee_reverse} would require classifiers to be trained and transferred to a device under test. TrustZone software binaries, for instance, are typically encrypted and authenticated during secure boot sequences on consumer devices. Moreover, TrustZone OSs are notoriously closed-source and closed-access, preventing users from installing custom TAs after deployment for acquiring ground-truth labels. A future research direction is to explore the efficacy of \emph{transfer learning} by training a model initially in an accessible white-box environment where samples can be reliably labelled before transposing it to black-box cases.  Although GlobalPlatform TEE APIs \emph{specify} a standard set of algorithms, one challenge with this approach is that their \emph{implementations} may differ during the transfer between white-box and black-box, OEM implementations.  While we evidenced that different library implementations may be classified under a single label in \S\ref{sec:algorithm_identification}, this remains untested in a transfer learning environment.

Lastly, we stress that our approach was evaluated on regular programs. Obfuscated binaries, particularly those that maliciously perturb HPCs to thwart PMU-based methods, can exhibit more complex behaviours that may pose generalisation challenges, which we defer to future research.

%% file: sections/conc.tex
\section{Conclusion}

This paper developed, implemented and evaluated a generic approach to black-box program function recognition. Extending work from related HPC research, we explored a side-channel approach in which micro-architectural events are analysed in bulk using a generic machine learning workflow. We then presented a three-part evaluation of this approach: \Circled{1} A preliminary study that examined classifying functions from widely used benchmarking suites and cryptographic libraries; \Circled{2} detecting several known, CVE-numbered vulnerabilities within OpenSSL; and \Circled{3} cryptographic function recognition within ARM TrustZone.  The approaches achieved 86.22\%--99.83\% accuracy depending on the target architecture and application. We showed how functions are recognisable in a relatively large, multi-class problem space. Furthermore, we demonstrated how OpenSSL lettered versions containing security vulnerabilities can be identified with high accuracy (0.889--0.998 $F_1$-score; 89.58\%--99.83\% accuracy). We then presented results from a further use case for recognising functions in a reference open-source ARM TrustZone TEE implementation with high accuracy (95.50\%--97.45\%) using a comprehensive range of GlobalPlatform TEE API functions. The results our work are broadly commensurate with alternative approaches, albeit with different adversarial models, including physical EM analysis for detecting cryptographic operations, malware detection, and OS detection (84--99\%) and static analysis and symbolic execution (84\%--100\%). Rather than supplanting these methods, we hope that our approach offers new thinking for software analysis using HPCs.

We posit that focussing on well-known side-channel vectors---cache accesses, timing differences, and branch predictions----are insufficient for engineering implementations which are resistant to the methods presented in this paper. Focus must directed to a wider range of micro-architectural events simultaneously---TLB misses, instruction retirements, clock cycles, pre-fetch events, and others---rather than prescribed events popularised in related work. We also re-emphasise best-practice guidelines for configuring PMUs to avoid exposing TEE side-channels to untrusted applications. Further work is needed to leverage more sophisticated attack scenarios, e.g.\ interpreted languages and transferring models \emph{between} consumer devices, but we present evidence that HPC-based function recognition is effective on today's major CPU architectures.